\documentclass{aastex63}

\usepackage{
graphicx,
color,
natbib
}

\newcommand{\nh}{\emph{New Horizons}\ }
\newcommand{\hst}{\emph{HST}\ }
\newcommand{\onebyone}{1$\times$1\ }
\newcommand{\fourbyfour}{4$\times$4\ }

\accepted{PASP, January 2020}
\shortauthors{Weaver et al.}

\begin{document}

\title{In-Flight Performance and Calibration of the LOng Range Reconnaissance Imager (LORRI) for the \nh Mission}

\correspondingauthor{H. A. Weaver}
\email{hal.weaver@jhuapl.edu}

\author[0000-0003-0951-7762]{H. A. Weaver}
\affil{Johns Hopkins University Applied Physics Laboratory,
11100 Johns Hopkins Road,
Laurel, MD 20723-6099, USA}

\author{A. F. Cheng}
\affil{Johns Hopkins University Applied Physics Laboratory,
11100 Johns Hopkins Road,
Laurel, MD 20723-6099, USA}

\author{F. Morgan}
\affil{Johns Hopkins University Applied Physics Laboratory,
11100 Johns Hopkins Road,
Laurel, MD 20723-6099, USA}

\author{H. W. Taylor}
\affil{Johns Hopkins University Applied Physics Laboratory,
11100 Johns Hopkins Road,
Laurel, MD 20723-6099, USA}

\author{S. J. Conard}
\affil{Johns Hopkins University Applied Physics Laboratory,
11100 Johns Hopkins Road,
Laurel, MD 20723-6099, USA}



\author{J. I. Nunez}
\affil{Johns Hopkins University Applied Physics Laboratory,
11100 Johns Hopkins Road,
Laurel, MD 20723-6099, USA}

\author{D. J. Rodgers}
\affil{Johns Hopkins University Applied Physics Laboratory,
11100 Johns Hopkins Road,
Laurel, MD 20723-6099, USA}

\author{T. R. Lauer}
\affil{NSF's National Optical-Infrared Astronomy Research Laboratory,
P. O. Box 26732,
Tucson, AZ 85726, USA}

\author{W. M. Owen}
\affil{Jet Propulsion Laboratory,
4800 Oak Grove Drive,
Pasadena, CA 91109, USA}

\author{J. R. Spencer}
\affil{Southwest Research Institute,
1050 Walnut Street, Suite 300,
Boulder, CO 80302, USA}

\author{O. Barnouin}
\affil{Johns Hopkins University Applied Physics Laboratory,
11100 Johns Hopkins Road,
Laurel, MD 20723-6099, USA}

\author{A. S. Rivkin}
\affil{Johns Hopkins University Applied Physics Laboratory,
11100 Johns Hopkins Road,
Laurel, MD 20723-6099, USA}




\author{C. B. Olkin}
\affil{Southwest Research Institute,
1050 Walnut Street, Suite 300,
Boulder, CO 80302, USA}


\author{S. A. Stern}
\affil{Southwest Research Institute,
1050 Walnut Street, Suite 300,
Boulder, CO 80302, USA}

\author{L. A. Young}
\affil{Southwest Research Institute,
1050 Walnut Street, Suite 300,
Boulder, CO 80302, USA}

\author{M. B. Tapley}
\affil{Southwest Research Institute,
6220 Culebra Road,
San Antonio, TX 78238, USA}

\author{M. Vincent}
\affil{Southwest Research Institute,
1050 Walnut Street, Suite 300,
Boulder, CO 80302, USA}




\begin{abstract}

The LOng Range Reconnaissance Imager (LORRI) is a 
panchromatic \mbox{(360--910 nm} for the wavelengths where the responsivity falls to 10\% of the peak value), narrow-angle 
\mbox{(field of view $=$ 0 \fdg 29),} 
high spatial resolution (pixel scale $=$ 1\farcs 02) visible light imager used on NASA's \nh (NH) mission for both
science observations and optical navigation.
Calibration observations began several months after the \nh launch on \mbox{2006 January 19}
and have been repeated approximately annually throughout the course of the mission,
which is ongoing.
This paper describes the in-flight LORRI calibration measurements, and the results derived
from our analysis of the calibration data.
LORRI has been remarkably stable over time with no detectable changes (at the $\sim$1\% level) in
sensitivity or optical performance since launch.
The point spread function (PSF) varies over the FOV but is well-characterized and stable, enabling
accurate deconvolution to recover the highest possible spatial resolution during observations
of resolved targets, especially when multiple, overlapping images are obtained.
By employing \mbox{$4 \times 4$} re-binning of the CCD pixels during read out, a special
spacecraft tracking mode, exposure times of $\sim$30~s, and co-addition of $\sim$100 images, 
LORRI can detect unresolved targets down to \mbox{$V \approx 22$} with a signal-to-noise ratio (SNR)
of $\sim$5.
LORRI images have an instantaneous dynamic range of $\sim$3500, which combined with exposure
time control ranging from 0~ms to 64,967~ms in 1~ms steps supports high resolution, high sensitivity imaging of 
planetary targets spanning heliocentric distances from Jupiter to deep in the Kuiper belt, enabling a wide variety of
scientific investigations.
We describe here how to transform LORRI images from raw (engineering) units into
scientific (calibrated) units for both resolved and unresolved targets.
Assuming that the wavelength variation of LORRI's sensitivity is accurately
described by the ground-based calibration, we estimate that LORRI's absolute
sensitivity is accurate to $\sim$2\% (1$\sigma$) for targets with solar-type
spectral energy distributions (SEDs). 
The accuracy of the absolute calibration for targets with other SEDs
should be comparably good when employing synthetic photometry techniques,
which we do when deriving LORRI's photometry keywords. 
We also describe various instrumental artifacts that could affect the interpretation of LORRI images
under some observing circumstances.

\end{abstract}

\keywords{Astronomical Instrumentation --- telescopes --- instrumentation: detectors --- methods: data analysis ---
space vehicles: instruments --- techniques: photometric --- }


\section{Introduction to LORRI} \label{sec:intro}
\begin{figure}[b!]
\includegraphics[keepaspectratio,width=\linewidth]{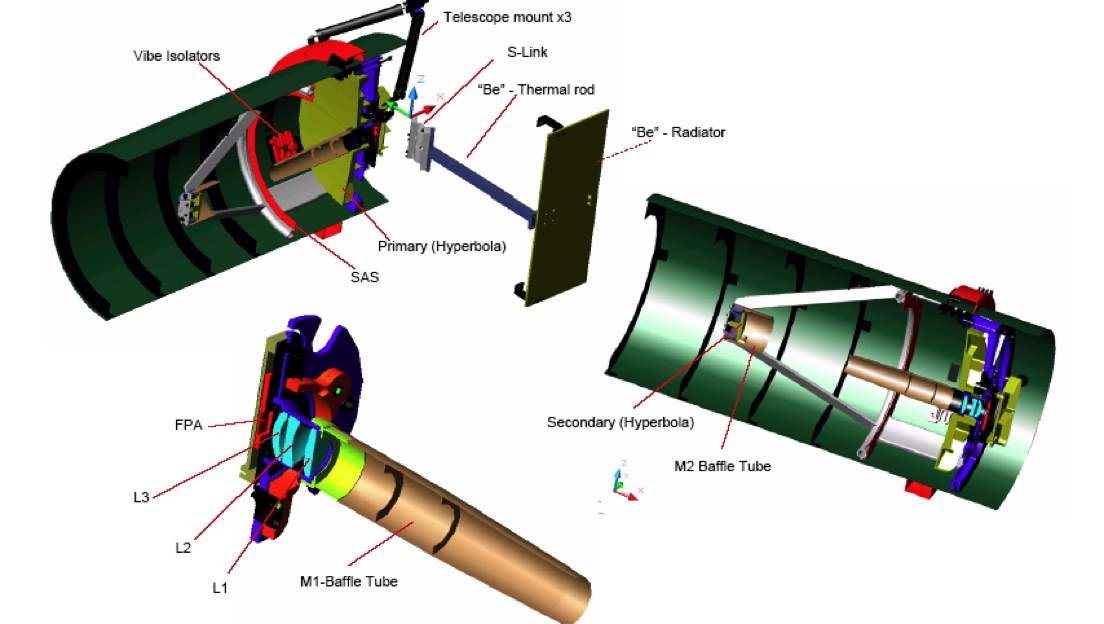}
\caption{Computer Aided Design (CAD) views of LORRI.
``M1'' refers to the primary mirror, ``M2'' refers to the secondary mirror,
``FPA'' refers to the focal plane assembly, ``Be'' refers to beryllium components, 
and ``SAS'' refers to the system aperture stop.
``L1'', ``L2'', and ``L3'' refer to the 3 separate lenses of the field-flattening lens assembly.
}
\label{fig:lorri_cadview}
\end{figure}

The LOng Range Reconnaissance Imager (LORRI) is a narrow angle (FOV=0.291\arcdeg), high spatial resolution (IFOV=1\farcs02),
Ritchey-Chr\'{e}tien telescope with a 20.8 cm diameter primary mirror, a focal length of 262 cm, and a three lens field-flattening
assembly (Figure~\ref{fig:lorri_cadview}).

The telescope, its lens assembly, and its baffle together are referred to as the Optical Telescope Assembly (OTA),
which was provided by SSG Precision Optronics, now L3 Harris SSG.
The telescope metering structure and its two mirrors are constructed from silicon carbide, providing an optical
design that maintains focus from \mbox{$+$50\arcdeg C} to \mbox{$-$120\arcdeg C} without the use of a focus mechanism
when mounted within the \nh spacecraft.
LORRI's only moving part is a once-open aperture cover mounted to the \nh spacecraft, which was opened on \mbox{2006 August 29,} 
after LORRI was allowed to outgas for decontamination purposes at a temperature of 
approximately $+$50\arcdeg~C for approximately 7 months.
After the door opened, the OTA mirrors have generally remained within several degrees of $-$70\arcdeg~C,
except when a heater is turned on to promote additional decontamination, when the OTA mirror temperatures rise to 
approximately $-$35\arcdeg~C.
Table~\ref{tab:decons} details how many times the 10~W decontamination heater has been activated
during the course of the mission.

\begin{deluxetable*}{lc}
\tablecaption{LORRI decontamination activities \label{tab:decons}}
\tablewidth{\linewidth}
\tablehead{
\colhead{Year} & \colhead{Number of decontamination activities} 
}
\startdata
2006		& 2$^{*}$ \\
2007		& 8 \\
2008		& 6 \\
2009		& 2 \\
2010		& 3 \\
2011		& 3 \\
2012		& 2 \\
2013		& 2 \\
2014		& 3 \\
2015		& 3 \\
2016		& 3 \\
2017		& 4 \\
2018		& 1 \\
2019		& 1 \\
\enddata
$^{*}$After the LORRI front aperture door was opened on 2006 August 29, there
were two additional decontamination activities that year.
From launch on 2006 \mbox{January 19} until the door was opened, 
LORRI was allowed to outgas for decontamination purposes at a temperature of 
approximately $+$50\arcdeg~C.
\tablecomments{A ``decontamination activity'' involves turning on the
10~W decontamination heater for at least 24~hr, and sometimes
for as long as $\sim$6 months.
In  2018-2019, we showed that the 10~W decontamination heater
can be turned off for at least 6 months without affecting LORRI's optical performance.
Thus, we plan to have fewer decontamination activities during post-2019 operations,
typically once per year.
}
\end{deluxetable*}

\begin{figure}[h!]
\includegraphics[keepaspectratio,width=\linewidth]{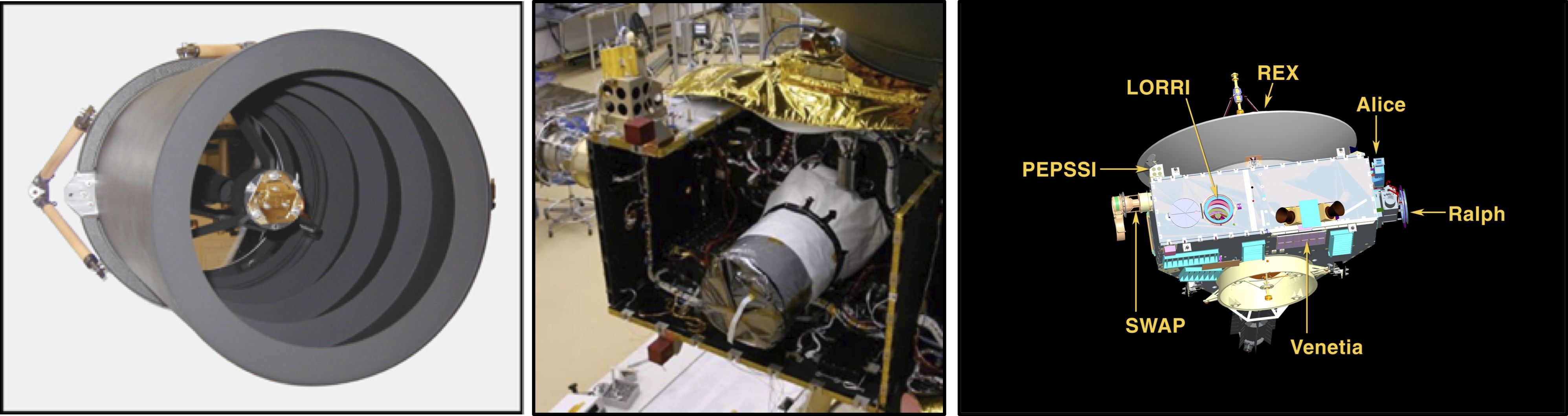}
\caption{The image on the left shows a completely assembled LORRI on a lab bench
during ground testing. The middle image shows where LORRI is mounted within
the \nh spacecraft. The image on the right is a drawing showing the locations of
all the instruments on the \nh spacecraft.
The diameter of the high gain antenna, which feeds the REX instrument, is 2.1~m.
Short descriptions of all the \nh instruments can be found in \citet{weaver:2008ssr}.
}
\label{fig:lorri_spacecraft}
\end{figure}

Figure~\ref{fig:lorri_spacecraft} shows the LORRI OTA during laboratory testing, 
LORRI's location within the \nh spacecraft, and the locations
of all the instruments on the \nh spacecraft.
The boresights of the three remote sensing instruments (LORRI, Ralph, and Alice) are approximately co-aligned.

A \mbox{$1024 \times 1024$} pixel (optically active region), back-thinned, backside-illuminated CCD detector
(model CCD~47-20 from e2v, now Teledyne-e2v) is located at the telescope focal plane  and is operated in standard frame-transfer mode.
The LORRI CCD incorporates anti-blooming technology to eliminate bleeding of the electrons along columns
when bright targets saturate (i.e., when the signal in a pixel exceeds the full-well capacity of $\sim$80,000 electrons).
The optically active area of the CCD has a ``midband'' anti-reflection coating that provides high
quantum efficiency across the visible portion of the spectrum, exceeding $\sim$90\% at
wavelengths near the peak efficiency (see further discussion later).
For the highest resolution observations, all optically active pixels are read out from the CCD (``1$\times$1'' format).
But the pixels can be also be re-binned by a factor of 4 in each direction (i.e., column and row directions)
during CCD readout (``4$\times$4'' format), which reduces the data volume by a factor of 16 and 
results in an effective pixel size of IFOV$=$4\farcs08.
During readout in either format, the analog signals are processed using correlated double-sampling and converted
to data numbers (DNs) using a 12-bit analog-to-digital converter, yielding a valid DN integer range from 0 to 4095.

In addition to the optically active pixels, four columns (one column in \fourbyfour format) in the 
non-illuminated region of the CCD are read out to provide a measurement of the bias level and dark current
for each image.
The bias level is $\sim$540~DN but varies linearly with the temperature of the focal plane electronics board.
The readout of the extra CCD columns enables accurate tracking of the bias level for each image.
Further discussion of how the calibration pipeline determines the bias level is provided below.

\begin{figure}[htb!]
\includegraphics[keepaspectratio,width=\linewidth]{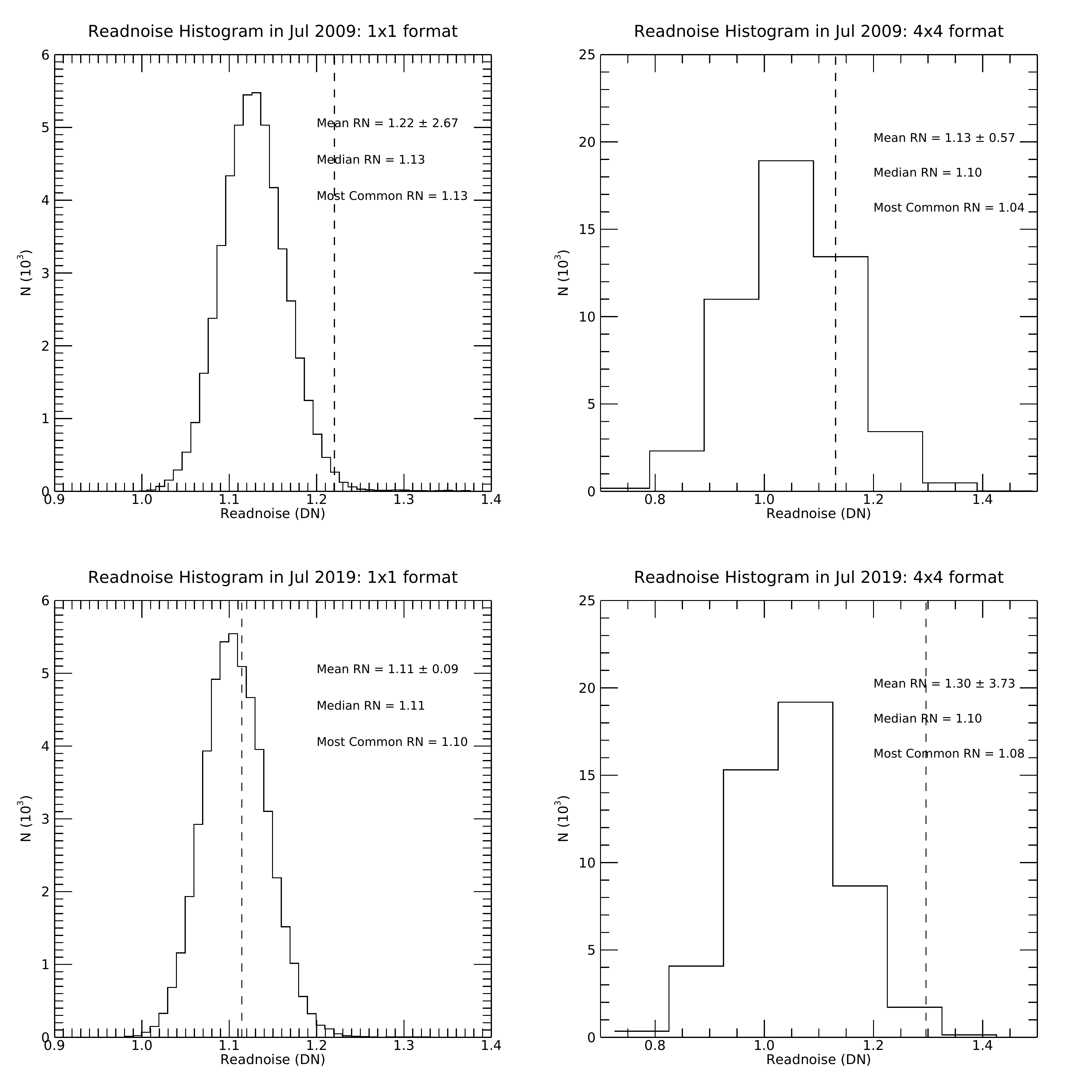}
\caption{These figures show in-flight electronics noise histograms for LORRI at two different epochs
separated by 10 years: July 2009 data are shown along the top row and July 2019 data are shown along the bottom row.
The figures on the left are for \onebyone CCD format, and the figures on the right are for \fourbyfour CCD format.
The mean (dashed vertical line), standard deviation, and median values are displayed for each histogram, but the most common
values (also displayed) should be more representative of the true electronics noise. In any case, there has been essentially
no change in the electronics noise over the entire New Horizons mission.}
\label{fig:rn_histograms}
\end{figure}

The electronics noise, which includes the CCD read noise, is $\sim$1.1~DN.
The electronics noise is monitored throughout the mission by differencing two bias images (i.e., with commanded exposure times of 0~ms), 
which are obtained during each annual checkout.
Following the technique outlined by \citet{janesick:2001}, we randomly sample \mbox{25 $\times$ 25} pixel subarrays
(\mbox{7 $\times$ 7} pixel subarrays for \fourbyfour images) of the difference image for 50,000 randomly selected
regions across the entire CCD to obtain a histogram of noise values.
Examples of noise histograms obtained both early in the mission (July 2009) and more recently (July 2019) are shown
in Figure ~\ref{fig:rn_histograms}.
Since the histograms are not completely symmetrical (the images have cosmic rays), neither the average value nor the median value
give the best representation of the noise.
Rather, we use an eyeball estimate of the peak in the histogram, after generating the histogram multiple times (since random numbers
are used to determine the box centers, the computer program produces a slightly different histogram each time it is run),
as our best estimate of the actual electronics noise.
The effective dynamic range for a single image is $\sim$3500, which is the largest DN value available (after accounting
for the bias level) divided by the noise.

\begin{figure}[htb!]
\includegraphics[keepaspectratio,width=\linewidth]{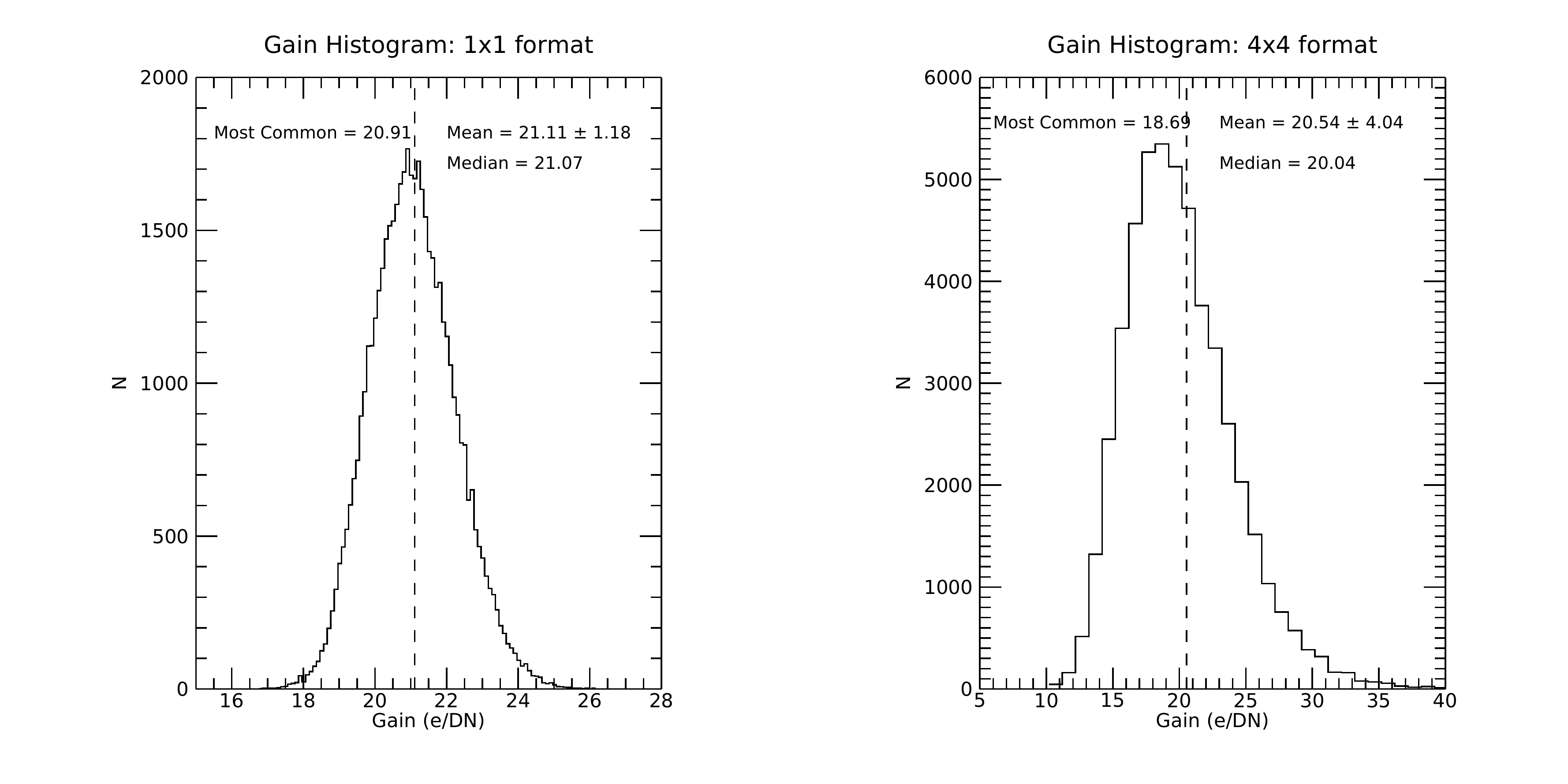}
\caption{These figures show gain histograms obtained during ground testing for each of the two CCD formats.
For both  \onebyone format (left) and \fourbyfour format (right), the means (dashed vertical lines), standard deviations, and median values are displayed for the histograms, but the most common values (also displayed) should be more representative of the true electronics gain.
As discussed in the text, we have adopted a gain of $21.0$~e~DN$^{-1}$ for \onebyone format and $19.4$~e~DN$^{-1}$  for \fourbyfour format.}
\label{fig:gain_histograms}
\end{figure}

The electronics gain, which is the scale factor that enables conversion from electrons detected to DN,
is slightly different for \onebyone versus \fourbyfour images.
Again, we follow the technique outlined by \citet{janesick:2001} to produce a gain histogram for each
of the two CCD formats using two well-exposed images with essentially constant illumination over the entire CCD.
In this case, however, we use differences of two flat field images taken from the post-environmental calibration conducted in September 2004,
prior to launch, to produce gain histograms because no in-flight flat fields could be obtained.
Examples of gain histograms are shown in Figure~\ref{fig:gain_histograms}, and they clearly show
that \fourbyfour images have a slightly higher gain than \onebyone images.
This same behavior has been seen in flight when observing the same targets in the two different CCD formats.
For example, in October 2008 when taking alternating \onebyone and \fourbyfour images of solar stray light
while the spacecraft rolled about the LORRI boresight at a fixed solar elongation angle of $\sim$13\arcdeg, the ratios of
the median signal rates was $\sim$$17.3$ (\fourbyfour / \onebyone) rather than the expected value of 16.
Similarly, the integrated signals of stars \mbox{(in DN s$^{-1}$)} appear to be $\sim$8\% larger in \fourbyfour images
compared to \onebyone images of the same fields.
We have adopted gain values of $21.0$~e~DN$^{-1}$ and $19.4$~e~DN$^{-1}$ for \onebyone and \fourbyfour images, respectively,
which are consistent with both the pre-launch and post-launch data.
With these gain values, the full well capacity of the CCD is nicely matched to the full range of the analog-to-digital converter.  

LORRI's response is linear to within 1\% over its entire dynamic range, as demonstrated during both ground
testing \citep{morgan:2005} and in-flight observations taken over a range of different exposure times.
Cooling of the CCD is provided by a beryllium radiator, which is painted white and is connected to the back end
of the CCD mount via an S-link and a conduction rod.
The radiator is exposed to space on the side of the spacecraft opposite that of the high gain antenna.
During a period of outgassing with the aperture door closed during the first 7 months of the \nh mission, 
the CCD temperature was warmed to $\sim$30\arcdeg~C.
Periodically throughout the mission (see Table~\ref{tab:decons}), a 10~W decontamination heater is used 
to warm the CCD to approximately $-$35\arcdeg~C to prevent accumulation of any outgassed contaminants.
Otherwise, the LORRI CCD temperature has been stable at approximately $-$80\arcdeg~C throughout the mission,
which is low enough that dark current is negligible for all 1$\times$1 format images.
Dark current is detectable in ``warm'' pixels during long exposures taken in 
\fourbyfour format. 
The location and number of warm pixels is variable, but is always $<$1\% of the total number of pixels.

As is the case for all imagers using CCDs, the charge transfer efficiency (CTE) is less than 100\%, which can result
in lower signal rates for targets located at high row numbers (i.e., farthest from the serial transfer output) and potential
degradation of the PSF.
We did not independently measure the CTE, but the manufacturer's specification sheet gives CTE values of
$0.999993$ for serial direction transfers (i.e., along the row direction) and $0.999999$ for parallel direction
transfers (i.e., along the column direction).
Measurements of stars in different regions of the CCD (described later) suggest no change in the CCD CTE values
over the course of the mission at the level of $\sim$1\%.

LORRI does not have a shutter. Whenever LORRI is active, the CCD is exposed to whatever scene is
in the FOV. The clocking of the CCD includes a ``frame scrub'', followed by exposure to the scene
 for the commanded integration time, followed by a ``frame transfer''  in which the CCD rows are sequentially
 transferred from the optically active area to the image storage region, followed by a readout of the image
 storage region to the downstream electronics, which convert the detected electrons to DNs.
 The digitized image is then transferred to the spacecraft's solid state recorder (SSR).
 The spacecraft's command and data handling (CDH) computer can either losslessly compress the full image,
 lossy compress the full image, or window and losslessly compress arbitrarily selected portions of the image.
 When commanded to do so, the CDH computer can send selected images from the SSR to the spacecraft's
 telecommunications hardware for downlink to the antennas of NASA's deep space network (DSN).
 For observations conducted in the Kuiper belt (i.e., when the heliocentric distance is $\geq$30~AU), 
 data downlink rates are typically $\sim$2~kbps when both of the redundant transmitters are used, or $\sim$1~kbps
 when only a single transmitter is used.
 
LORRI does not have any color filters.
Instead, LORRI maximizes sensitivity by providing panchromatic imaging over a wide bandpass:
\mbox{435--870 nm} at the 50\% throughput points and \mbox{360--910 nm} at the 10\% throughput points.
Images can be taken at a rate of up to once per second, or at any commanded cadence longer than that.
Manual exposure times can be commanded from 0~ms to 64,967~ms in 1~ms increments.\footnote{New LORRI flight 
software was uploaded in July 2019 that enabled exposure times up to 64,967~ms. 
Prior to that, the longest available exposure time was 29,967~ms.
Observations taken in September 2019 have
verified nominal performance with exposure times of 64,967~ms.} 
We recently (July 2019) discovered an error in the exposure times reported
by the LORRI flight software. As a consequence, the actual exposure time is $\sim$0.6~ms longer
than reported. This error is insignificant for most LORRI images, but it is important for many
of the LORRI images taken during the Jupiter flyby, when the commanded exposure times
were generally $\leq$10~ms.
We note also that using the actual exposure time is important for correcting the smear produced during the frame scrub
and frame transfer process, as described later in more detail.
The LORRI data will be re-processed to account
for this error in the exposure times for a future delivery to the Planetary Data System archive.

A flexible autoexposure mode is available whenever the scene being imaged has unknown intensities, 
but autoexposure can only be used for exposure times $\leq$967~ms.
Although we exercise the autoexposure capability at every annual checkout,
we have generally preferred to use two different \emph{manual} exposure times that span the dynamic range of interest, rather than relying
on autoexposure mode for in-flight observations of targets with unknown, or poorly characterized, brightnesses.

A ``trigger'' mode is also available that enables LORRI to determine autonomously when a target has entered
the FOV, via analysis of a 32-bin histogram of the image, and then send images to the spacecraft for a 
specified duration after the trigger condition has been satisfied.
Trigger mode was developed in 2011 and required uplinking to the spacecraft new LORRI flight software.
The primary potential use of trigger mode was to enable taking many (tens to hundreds) \onebyone images
during the Pluto flyby as the LORRI FOV was scanned across Pluto and Charon and their very large pointing
error ellipses without  filling up the solid state recorder with many dozens of blank sky images.
Although trigger mode was tested extensively on the ground using the LORRI engineering model,
and trigger mode also worked successfully during an in-flight test scanning the LORRI FOV across a bright star,
no in-flight test of an \emph{extended} target could be identified that mimicked the conditions of the Pluto flyby.
For that reason, the \nh Project decided \emph{not} to use trigger mode during the Pluto flyby.
Trigger mode was also not  used during the \nh spacecraft flyby of the 
Kuiper belt object 2014~MU$_{69}$ (hereafter ``MU69'') because that target was so faint that the necessary trigger condition
could not be satisfied with the coarse (32 bins) image histogram available.

The \nh spacecraft has two main guidance and control (GNC) states: either spinning at $\sim$5~rpm
about an axis approximately aligned with the high gain antenna and the LORRI boresight, or 3-axis stabilized mode.
When operating in 3-axis mode, which is usually the case during science
observations for the remote sensing instruments, the \nh spacecraft fires hydrazine thrusters to
maintain the pointing within a specified attitude ``deadband''.
(Owing to power constraints, \nh doesn't have reaction wheels to stabilize the pointing.)
The deadbands used during LORRI \onebyone observations are either $\pm$51\farcs6 or $\pm$103\arcsec,
with thruster firings approximately every 3~s and every 6~s, respectively.
Typical pointing drift rates during 3-axis observations are $\sim$5\arcsec~s$^{-1}$;
typical \onebyone exposure times are 0.10$-$0.15~s during science observations to limit
pointing smear to $\leq$1~pixel.

\begin{figure}[htb!]
\includegraphics[keepaspectratio,width=\linewidth]{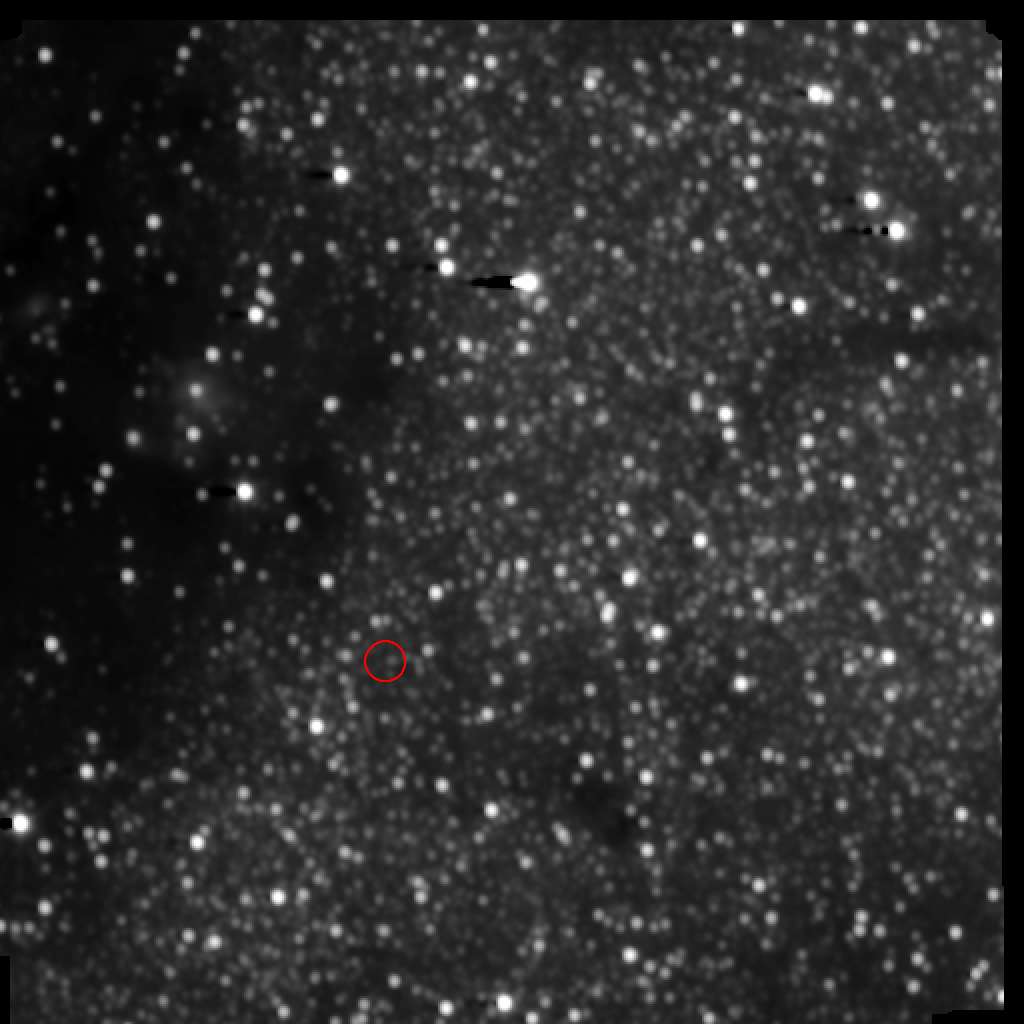}
\caption{This deep LORRI image of the MU69 background field was produced by combining 96 individual
30~s images taken on \mbox{2017 September 21.}
Celestial north is up, and east points to the left.
The pixels are subsampled by a factor of two (using sinc function interpolation) to remove the pixelated appearance
of the raw \fourbyfour image.
A hyperbolic sine ($\sinh$) intensity stretch (similar to logarithmic but better behaved near 0) ranging from $-$1 to 3500~DN is
used to show the full dynamic range of the image.
The brightest stars are saturated and have black tails due to amplifier undershoot.
The red circle is centered on the predicted location of MU69, but MU69 was fainter than LORRI's sensitivity
limit, which is estimated to be $V$$\approx$22 in this composite.
This composite image was used as a template that could be subtracted from LORRI images of MU69
taken in late-2018, thereby removing nearby stars and enabling the detection of MU69 for optical
navigation and deep searches for satellites and dust in the vicinity of MU69. 
}
\label{fig:lorri_deep}
\end{figure}

Shortly after launch, a new 3-axis pointing mode, called ``relative control mode'' (RCM),
was developed to enable long duration exposures in \fourbyfour format.
In RCM, the thrusters are fired $\sim$80 times per minute to keep the
LORRI boresight at a fixed inertial pointing location to an accuracy of $\pm$2\arcsec (1$\sigma$)
for exposure times up to $\sim$65~s.
RCM has enabled LORRI deep imaging observations that can reach \mbox{$V \approx 20.2$}
with \mbox{SNR$=$5} in a single 65~s image for targets with solar-type spectral energy distributions (SEDs)
in relatively sparse fields.
Figure~\ref{fig:lorri_deep} shows a composite LORRI \fourbyfour image of the MU69 background field produced by combining
96 individual 30~s exposures, which reaches \mbox{$V \approx 22$} (SNR$=$5) in the regions 
with relatively few background stars.

\begin{figure}[h!]
\includegraphics[keepaspectratio,width=\linewidth]{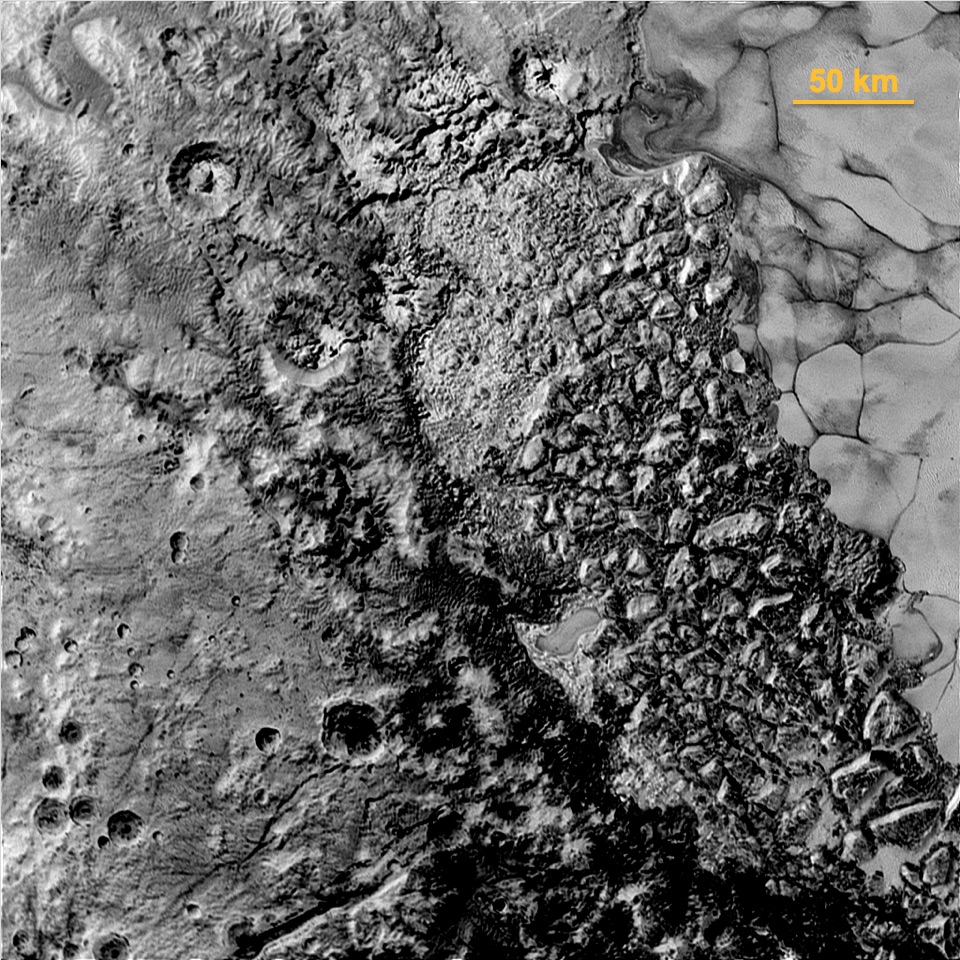}
\caption{This is a deconvolved LORRI image of the al-Adrisi mountains on Pluto taken on \mbox{2015 July 14}
when the spacecraft was 78,600~km from the surface \mbox{(0.390 km pixel$^{-1}$).}
The exposure time was 0.15~s, which is in the typical range used for science observations
when LORRI is the prime instrument, with its boresight fixed relative to the target.
The region to the right of the frame is the northwest corner of Sputnik Planitia (SP), a giant ice sheet
composed primarily of N$_{2}$ that displays a geometrical pattern attributed to rising (near the center
of a pattern) and falling (at the boundaries of the patterns) icy material heated from below.
The mountains are the blocky structures at the edge of SP and are thought to be giant water icebergs
floating in the nitrogen ice. Many other interesting geological features are evident in this image,
as discussed in \citet{moore:2016science}. 
}
\label{fig:al_adrisi}
\end{figure}

For most LORRI science and calibration observations, the boresight is pointed at the target of interest,  
the CCD is exposed for the commanded duration (typically 100-150~ms), and the CCD is read out 
after the exposure is finished.
In this case, the exposure times are long enough to provide excellent signal-to-noise ratios (SNRs), 
but they are short enough to provide high resolution imaging performance without significant degradation 
as the target drifts within the pointing deadband.
\mbox{Figure~\ref{fig:al_adrisi} }shows an example of this type of observation taken during the approach to 
Pluto in July 2015.
Deconvolution (see later discussion), which restores signal to the core that was originally spread into the
PSF wings, was applied to sharpen the image and achieve nearly diffraction-limited resolution.

\begin{figure}[h!]
\includegraphics[keepaspectratio,width=\linewidth]{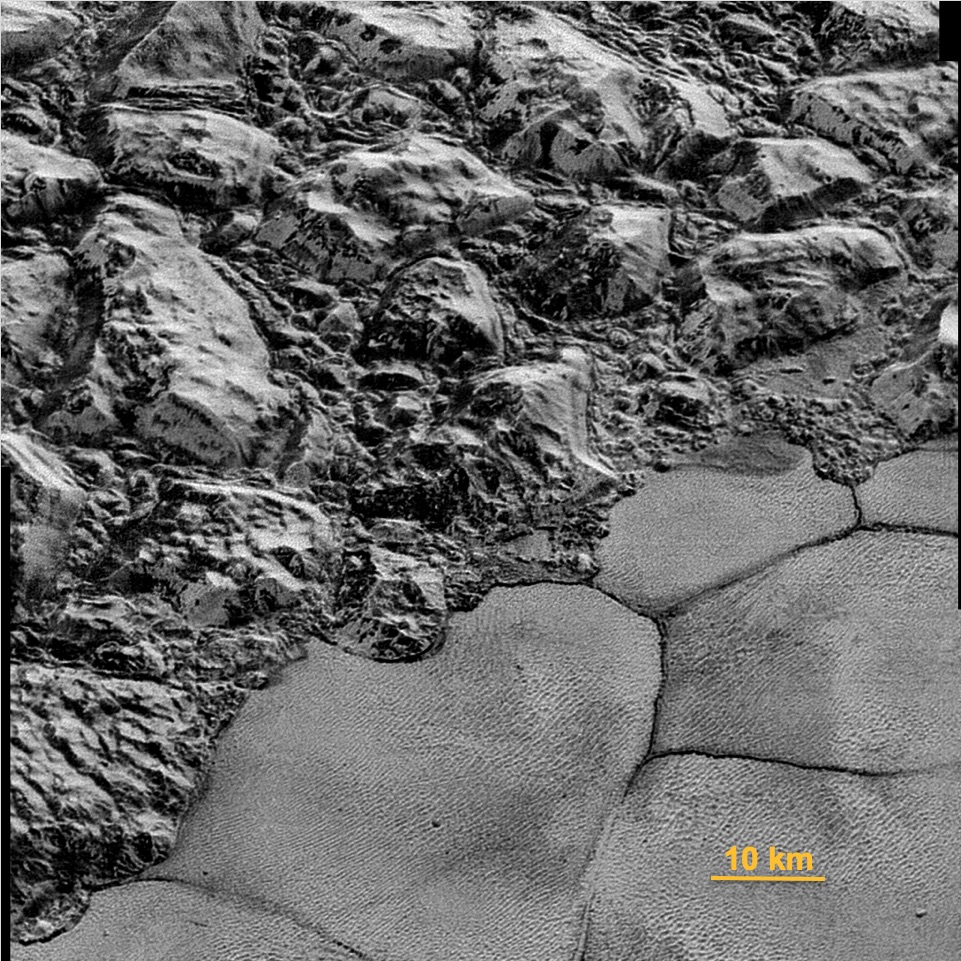}
\caption{This is a deconvolved LORRI image showing even finer detail on Pluto's surface, in this case the border between 
the Sputnik Planitia ice sheet (the smooth geometrical patterns below and to the right) and
 the chaotic terrain called the al-Adrisi mountain range. This image was taken on \mbox{2015 July 14}
when the spacecraft was 17,000~km from the surface \mbox{(0.084 km pixel$^{-1}$).}
The LORRI images were taken while scanning the MVIC imager across Pluto at \mbox{1000 $\mu$rad s$^{-1}$},
which required using LORRI exposure times of only 0.010~s to minimize pointing smear during the scan.
Although LORRI uses a standard frame transfer CCD, while MVIC uses a CCD in time delay 
integration (TDI) mode (``push broom mode''), LORRI's high sensitivity enabled its use 
during scanning observations.
}
\label{fig:chaos}
\end{figure}

However, during flyby encounters LORRI images are also frequently taken while the spacecraft is scanning the Ralph
instrument across the target and its error ellipse.
This mode enables simultaneous LORRI and Ralph observations, but the LORRI exposure times must
be much shorter to prevent excessive smear during the scan.
Typically LORRI images are taken either every second, or every 3 seconds, during such scans,
and the overlapping fields of view enable co-adding the images to recover better SNR.
In addition, deconvolution techniques can be employed to recover better optical resolution,
approaching that of the equivalent non-scanned observations.
Figure~\ref{fig:chaos} shows an example of a composite LORRI image obtained during a Ralph-MVIC
scan across the middle of Pluto near the time of closest approach.
This particular scan produced the highest resolution imaging of Pluto obtained during the flyby.

A detailed description of LORRI and its use on \nh can be found in \citet{cheng:2008ssr}.
Engineering details on the construction of LORRI can be found in \citet{conard:2005}.
The ground-based calibration of LORRI is described in \citet{morgan:2005}.
LORRI's performance through the Jupiter flyby is described in \citet{noble:2009},
and LORRI's performance during the Pluto flyby is described in \citet{conard:2017}.
LORRI's in-flight straylight performance is described in \citet{cheng:2010}, and
a detailed comparison of LORRI's optical design and its actual performance is
described in \citet{mcmichael:2012}.
\citet{zemcov:nature} describes attempts to measure the extragalactic background
optical light with LORRI, and \citet{zemcov:pasp} describe LORRI's potential
future uses for various astrophysical objectives.
Below we focus on the results from the in-flight LORRI calibration observations, 
their trending with time over the entire \nh mission to date,
and how the calibration results are used to derive the various photometry keywords needed to transform LORRI
images from engineering units to absolutely calibrated scientific units.

\section{Calibration Steps} \label{sec:cal}
Before discussing the in-flight LORRI calibration program and its results, we first describe the steps in the LORRI
calibration pipeline, which is the software that removes instrumental artifacts in the raw images and creates 
the calibrated images that can be used for scientific analysis.
The calibration pipeline makes use of reference files that are derived from either ground or flight calibration measurements,
as described further below.
Critically, the pipeline populates various photometry FITS header keywords based on the results
from LORRI observations of absolute calibration standard stars, as discussed in Section \ref{sec:obs}.

\subsection{Bias and Dark Subtraction} \label{subsec:bias}
After the LORRI aperture door was opened on 2006 August~29, the CCD temperature
has typically been within a few degrees of $-$81\arcdeg~C.
At this temperature, the CCD dark current is usually negligible for most LORRI observations. 
The process used to subtract the CCD electronic offset signal (which is called the ``bias'' level and is used to prevent sending
negative analog signals to the analog-to-digital converter) automatically includes any dark current that may be present.
From a series of long exposure time images (up to 64.967~s) taken during a test in July~2019 when
the CCD temperature was \mbox{$-$81\arcdeg C,} we determined
an \emph{upper limit} for the dark current in \onebyone pixels of \mbox{0.0019 DN s$^{-1}$ pixel$^{-1}$,} 
corresponding to \mbox{0.040 e s$^{-1}$ pixel$^{-1}$} at the nominal gain of \mbox{21 e pixel$^{-1}$},
which is approximately 10 times larger than the dark current calculated using data from the CCD
manufacturer's specification sheet.
During long exposure \mbox{($\geq$10 s)} \fourbyfour images obtained during flight, ``warm'' pixels are clearly detected
with elevated dark currents (up to a few DN in magnitude, after subtraction of the median bias plus dark level). 
The locations of these ``warm'' pixels vary over time, but they never comprise more than 1\% of the total
number of pixels in the image. 

The bias level is a function of the temperature of the focal plane electronics (FPE) board.
From the ground calibration data, we derived that the bias level for \onebyone images is given by:
\begin{equation}
\mathrm{bias} = 503.968 \,  (\pm 0.549) + 1.239 \, (\pm 0.015) \, \ast \, \mathrm{T_{board}}
\label{eqn:bias}
\end{equation}
where ``bias'' is the bias level in DN and T$_{\mathrm{board}}$ is the FPE board temperature.
For \fourbyfour images the bias level is typically $\sim$4~DN larger.
LORRI in-flight images have bias levels that are generally consistent with this formula.
After the LORRI aperture door was opened on 2006 August~29, the FPE board temperature
has typically been in the range 23\arcdeg~C to 29\arcdeg~C.

We have recently (in 2019) noticed that the bias level may also display a ``start-up'' feature, whereby the
bias level starts at a slightly higher value and gradually plateaus to its expected value over the
course of several minutes.
This behavior is still being characterized, but the magnitude of the effect is at the sub-DN level, which
is inconsequential for most \nh science applications.
We do, however, want to understand this effect better because it will likely affect LORRI's ability to
accurately measure the extragalactic background light \citep{zemcov:nature}.

As implemented in the LORRI pipeline, the bias$+$dark subtraction is a two-step process. 
First, the overall bias$+$dark level for an image is subtracted from the raw (``Level~1'')
image. Specifically, for \onebyone images (i.e., Level~1 files having \mbox{$1028 \times1024$ pixels)} the median intensity of the pixels in
columns 1024$-$1027 of the Level 1 image (corresponding to columns 1032$-$1035 of the CCD, which are in a portion of
the CCD not directly illuminated) is subtracted from each pixel in the Level~1 image.
For \fourbyfour images (i.e., Level~1 files having  \mbox{$257 \times 256$ pixels),} the median intensity of pixels in column 257 
(which is the re-binned version of CCD columns 1032$-$1035 in the \onebyone image) is subtracted from 
each pixel in the Level~1 image.
Since the bias columns also accumulate CCD dark current, the median dark current is also removed during
this step.

The simple median of the pixels in the inactive region may not provide the best estimate of the global \mbox{bias$+$dark} level,
if the distribution of pixel values is distorted by multiple cosmic rays, or other artifacts.
We have recently investigated an alternative approach, using fits to the well-defined peak in the histogram 
of the pixel values as the bias level estimate.
Although this technique may indeed provide a better estimate of the true bias level, we have found that the differences with
the median values are typically $\leq$0.3~DN, which is smaller than the typical electronics noise level ($\sim$1.1~DN).
We have decided to stick with the simple median for the near-term because this approach is more robust (i.e., not subject
to subtle computational issues) and easier to implement.
However, we may eventually employ a different technique to enable achieving the highest
possible sensitivity for the most demanding observations.

\begin{figure}[htb!]
\includegraphics[keepaspectratio,width=\linewidth]{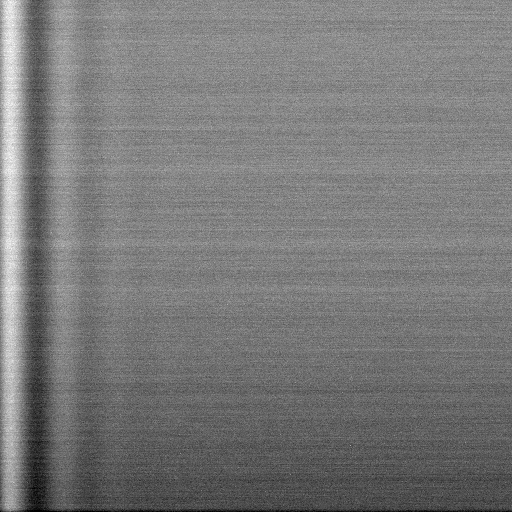}
\caption{This ``delta-bias'' image, which shows the pixel-to-pixel variations in the bias level, was created by 
combining 100 \onebyone images, each with an exposure time of 0~ms,
taken before the aperture door was opened.
The image is displayed using a linear intensity stretch ranging from $-$1 to $+$1~DN.}
\label{fig:delta_bias}
\end{figure}

After this global bias$+$dark level is subtracted, a ``delta-bias'' (sometimes called a ``superbias'') image is then subtracted from the
image. 
The delta-bias image captures the pixel-to-pixel variations in the bias and was created by averaging 100 \onebyone bias frames (i.e., images
with exposure times of 0~s) taken in-flight on \mbox{2006 July 30} with the main aperture door closed, 
as part of a LORRI in-flight calibration activity (Figure~\ref{fig:delta_bias}).
The delta-bias image for \fourbyfour format images was obtained by simple re-binning of the average \onebyone delta-bias image.
Although \fourbyfour bias images were obtained in April and May 2006, those were contaminated by light passing through the 
small circular window ($\sim$2.54~cm in diameter) in the main aperture door.
This window was placed in the aperture door to permit LORRI to have some imaging capability in case the main aperture door
didn't open.
The light passing through this small window was too weak to affect the \onebyone bias images, but could be faintly seen
in the \fourbyfour bias images.

Bias images in both \onebyone and \fourbyfour format have been taken at least once per year during the entire \nh mission.
No systematic change in the bias images has ever been seen.
LORRI images (both bias images and regular images) sometimes exhibit a small even-odd column signal offsets (``jail bars''), 
but the effect is generally
$\leq$0.5~DN in magnitude, which is smaller than the typical electronics noise ($\sim$1.1~DN).
We have not been able to correlate the appearance of the jail bars with any particular instrument parameter 
(e.g., there is no correlation with the temperatures of the CCD or the electronics). 
But if they are present, the jail bars are usually present in an entire set of images taken near the same time.
In that case, the jail bars can be removed fairly easily with standard image processing techniques (e.g., by creating a template image
with the jail bar pattern and subtracting that from the original image).

Individual LORRI images in both formats also display a low-level ($\leq$1~DN) horizontal (i.e., row-based) pattern
that is apparently associated with the injection of non-random noise during the CCD frame transfer process.
This ``horizontal striping'' is always present but is usually only evident when large portions of the image have
little or no signal (e.g., when observing sparse star fields).
The stripes change locations from frame-to-frame producing a non-random noise
pattern across each image.
This pattern can be removed from an individual image with standard image processing techniques
(e.g., by subtracting a median smoothed, or robust average, level from each row). 

The electronics noise is measured by subtracting one bias frame from another and examining the residuals.
This technique has been applied throughout the mission, and the results demonstrate that the electronics noise has
remained virtually the same for the entire mission ($\sim$1.1~DN) for both CCD formats.

\subsection{Desmear} \label{subsec:desmear}

As already mentioned above, LORRI does not have a shutter.
Thus, the target being observed illuminates the active region of the CCD whenever LORRI
is pointed at the scene. 
In particular, the CCD continues to record the scene as the ``scrub'' is performed (before the nominal
start of the exposure) and as charge is transferred from the active
portion to the storage area (after the nominal end of the exposure).
Both of these processes result in a smearing of the observed scene along a CCD column.

The smear process can be visualized as pulling a sheet of light sensitive film, whose size is exactly the size of the CCD's active area, 
uniformly across a scene whose size is also exactly the size of the CCD's sensitive area. 
The flush begins as the film is pulled across the scene starting from the top of the CCD (i.e., starting at the highest CCD row) 
until the film exactly covers the full scene. 
The total flush time is the duration of that process, and the rate of motion, which is also the row transfer time, 
determines how much signal from pixels at \emph{higher} rows is transferred to any particular pixel in the same column. 
After the flush process is completed, the film stays in place for the commanded exposure time, and each pixel accumulates signal 
only from the scene imaged at that pixel. 
After the exposure is completed, the film is pulled in the same direction as before, but this time each pixel will accumulate signal from the scene in rows 
below that pixel as the film is moved to the image storage region of the CCD, which is also completely outside the illuminated scene. 
The total transfer time is the duration of that process, and the rate of rate of motion during the transfer determines how 
much signal from pixels at \emph{lower} rows is transferred to any particular pixel in the same column. 
All of the pixels in the CCD storage region are then transferred row-by-row to the CCD serial readout register, 
where the analog signals are amplified and then digitized.

For images with no over-exposed pixels and that have fixed pointing during the exposure,
the smear can be essentially completely removed using the algorithm described in \citet{cheng:2008ssr}.
Once a pixel is over-exposed, however, information on its true intrinsic level is lost and the smear correction will
be incomplete for that column of the image.
As the de-smear algorithm works on a column by column basis, any significant motion of the camera
during the exposure in the row direction (i.e., perpendicular to the columns), as can occur when LORRI is used in ride-along mode with
the \nh MVIC or LEISA instruments, may also limit the accuracy of the correction.
Of course, the de-smear correction cannot remove the extra shot-noise associated with the smeared light.
For long exposures this is only a modest effect,
but for images of extended targets with short exposure times (i.e., similar to, or smaller than, 
the total frame transfer time of $\sim$12~ms),
the signals generated during the scrub and transfer processes are comparable to, or even larger than, the signal levels
accumulated during the nominal exposure time.
In this case, the shot noise from the smear signal produces significant degradation of the SNR.
For example, when the signal accumulated during the exposure time is comparable to
that accumulated during the scrub and transfer, the SNR is reduced by approximately $\sqrt{2}$
compared to the case when frame transfer smear is negligible.

The de-smear algorithm described in \citet{cheng:2008ssr} is complex and computationally intensive.
Given the sparse nature of the ``error matrix'' used in that formulation, we identified a technique
for speeding up the calculations by a factor of $\sim$10 compared to performing the matrix multiplications
given in \citet{cheng:2008ssr}.
This faster technique, which produces results essentially identical to those using the full
matrix multiplication, has been used in the LORRI data pipeline since 2014.

We have recently investigated two alternative formulations for the desmear step, which
are simpler and easier to implement than the algorithm given in \citet{cheng:2008ssr}.
One of the algorithms \citep{owen:2019} adopts approximations to accelerate the
computation by avoiding matrix operations.
Another algorithm, discussed further below, is exact but involves a matrix inversion for each image.
The new algorithms are still being evaluated and must be tested extensively before either can
be used in the LORRI calibration pipeline.

\subsubsection{An Exact Simple De-smear Algorithm}

We are presently investigating an alternative algorithm for the de-smear step, which
is simpler and easier to implement than the algorithm given in \citet{cheng:2008ssr}.
As with the \citet{cheng:2008ssr} approach,
the new algorithm is applied on a per column basis,
since the smeared charge associated with any pixel stays within its column.
The solution is derived with a single matrix multiplication that
transforms the column of data values as generated by LORRI to
a column of pixel values corrected for charge-transfer smear.

For any column of pixels in the LORRI CCD,
neglecting dark current, the rate at which signal is accumulated in a LORRI CCD pixel within that column can be written as:

\begin{equation}
\mathrm{
\frac{S_{k}}{t_{exp}} = \sum_{i,i>k}^{n} R_{i} \, I_{i} \, \frac{t_{scrub}}{t_{exp}} + \sum_{i,i<k}^{n} R_{i} \, I_{i} \, \frac{t_{transfer}}{t_{\mathrm{exp}}} + 
 R_{k} \, I_{k}
}
\label{eqn:smear}
\end{equation}

\noindent where:\\

\noindent $\mathrm{S_{k}}$ is the measured signal in a pixel in row k (electrons, or ``e'')\\
$\mathrm{R_{x}}$ is the intrinsic responsivity of pixel x ([e s$^{-1}$ pixel$^{-1}$] / [photons s$^{-1}$ pixel$^{-1}$])\\
$\mathrm{I_{x}}$ is the input photon flux at pixel x (photons s$^{-1}$ pixel$^{-1}$)\\
$\mathrm{t_{exp}}$ is the total exposure time (s), which is the commanded time plus 0.6~ms\\
$\mathrm{t_{scrub}}$ is the CCD row transfer time during the frame scrub (s)\\
$\mathrm{t_{transfer}}$ is the CCD row transfer time during the frame transfer (s)\\
$\mathrm{n}$ is the number of rows in the image (1024 for \onebyone images and 256 for \fourbyfour images)\\

The first term on the right-hand side of the equation represents the smear contribution to the observed signal
from pixels in the same column, but at \emph{larger} row numbers compared to the pixel of interest, and is accumulated
during the frame scrub process.
The second term on the right-hand side of the equation represents the smear contribution to the observed signal
from pixels in the same column, but at \emph{smaller} row numbers compared to the pixel of interest, and is accumulated
during the frame transfer process.
The third term is the actual desired quantity, the signal accumulated at the pixel of interest during the commanded exposure time.

Equation~\ref{eqn:smear} can be recast as a matrix equation:

\begin{eqnarray}
\mathrm{
D_{i} = \sum_{j=1}^{n} g_{ij} \, F_{j} } \\
\rightarrow \mathbf{D} = \mathbf{G} \, \mathbf{F}
\label{eqn:smatrix}
\end{eqnarray}

\noindent where in any CCD column:\\

\noindent $\mathrm{D_{i}}$ is the detected signal (including smear) at row i (DN)\\
$\mathrm{F_{j}}$ is the true signal (not including smear) at row j (DN)\\
$\mathrm{g_{ij}}$ are coefficients determined as described below\\

The coefficients of $\mathbf{G}$ are analogous to, but not identical to, the $\epsilon$ coefficients in
 the smear formulation of \citet{cheng:2008ssr}.
For LORRI, the frame scrub takes a total of 12.15~ms, which means each row transfer ($\mathrm{t_{scrub}}$ above) 
takes 0.0119~ms for \onebyone images and 0.0474~ms for \fourbyfour images.
The frame transfer takes 11.12~ms, which means each row transfer ($\mathrm{t_{transfer}}$ above) takes 0.0109~ms for \onebyone images
and 0.0434~ms for \fourbyfour images.
Note that \mbox{$\mathrm{g_{ij} = 1}$} when \mbox{$\mathrm{j=i}$,}
\mbox{$\mathrm{g_{ij} = t_{scrub}} / \mathrm{t_{exp}}$} when \mbox{$\mathrm{j > i}$,} and
\mbox{$\mathrm{g_{ij} = t_{transfer}} / \mathrm{t_{exp}}$} when \mbox{$\mathrm{j < i}$.}

In summary, the measured count rates for the pixels in a column (represented by the column matrix $\mathbf{D}$) can be represented by the 
matrix multiplication of the Òsmear matrixÓ and a column matrix of the actual count rates 
(i.e., $\mathbf{F}$, the count rates after the smear contributions have been removed). 
Thus, the desmear problem reduces to finding the inverse of the smear matrix:

\begin{equation}
\mathbf{F} = \mathbf{G^{-1}}  \, \mathbf{D}
\label{eqn:desmear}
\end{equation}

\noindent Note that the matrix inversion need only be performed once for each image
because the smear matrix is identical for each column in the image.
In fact, the smear matrix depends only on the ratios 
\mbox{$\mathrm{t_{scrub}} / \mathrm{t_{exp}}$} and
\mbox{$\mathrm{t_{transfer}} / \mathrm{t_{exp}}$.}
In principle, the matrix inversion could be performed in advance for each exposure time employed,
in which case the matrix inversion within the calibration pipeline could be replaced by
a simple lookup of the appropriate inverted matrix, which could be stored in a reference file 
directory (e.g., similar to what is done for the delta-bias and flat field).

\subsection{Flat-Field} \label{subsec:flat}
Flat-fielding refers to the process of removing the pixel-to-pixel sensitivity variations in the image. An exposure obtained
by illuminating the LORRI aperture uniformly with light is called a flat-field image. During ground calibration testing,
flat-fields were obtained by using an integrating sphere to provide uniform illumination \citep{morgan:2005}. The light source was a xenon
arc lamp with a spectrum similar to that of the Sun. The absolute intensity of the input illumination was measured using a
calibrated photodiode. For the panchromatic case, which is the one most relevant for flat-fielding LORRI images, the
light from the xenon lamp was unfiltered. Flat-field images were also obtained by passing the light through bandpass
filters centered at five different wavelengths spanning the range over which LORRI is sensitive, prior to injection into the
reference sphere, to estimate the sensitivity of the flat-fields to the spectral distribution of the source. The spatial
patterns in the flat-field images change significantly with wavelength. However, the variation in panchromatic
flat-fields caused by differences in the spectral distribution of the illumination source are much less significant.
Indeed, panchromatic flat-field images produced using a tungsten lamp were virtually indistinguishable from those
produced by the xenon lamp. Flat-fields were obtained at four different sets of thermal environments (at standard laboratory
room temperature, and at the lowest, nominal, and highest temperatures predicted for in-flight conditions), but no
significant variations in the flat-field images were detected.

\begin{figure}[htb!]
\includegraphics[keepaspectratio,width=\linewidth]{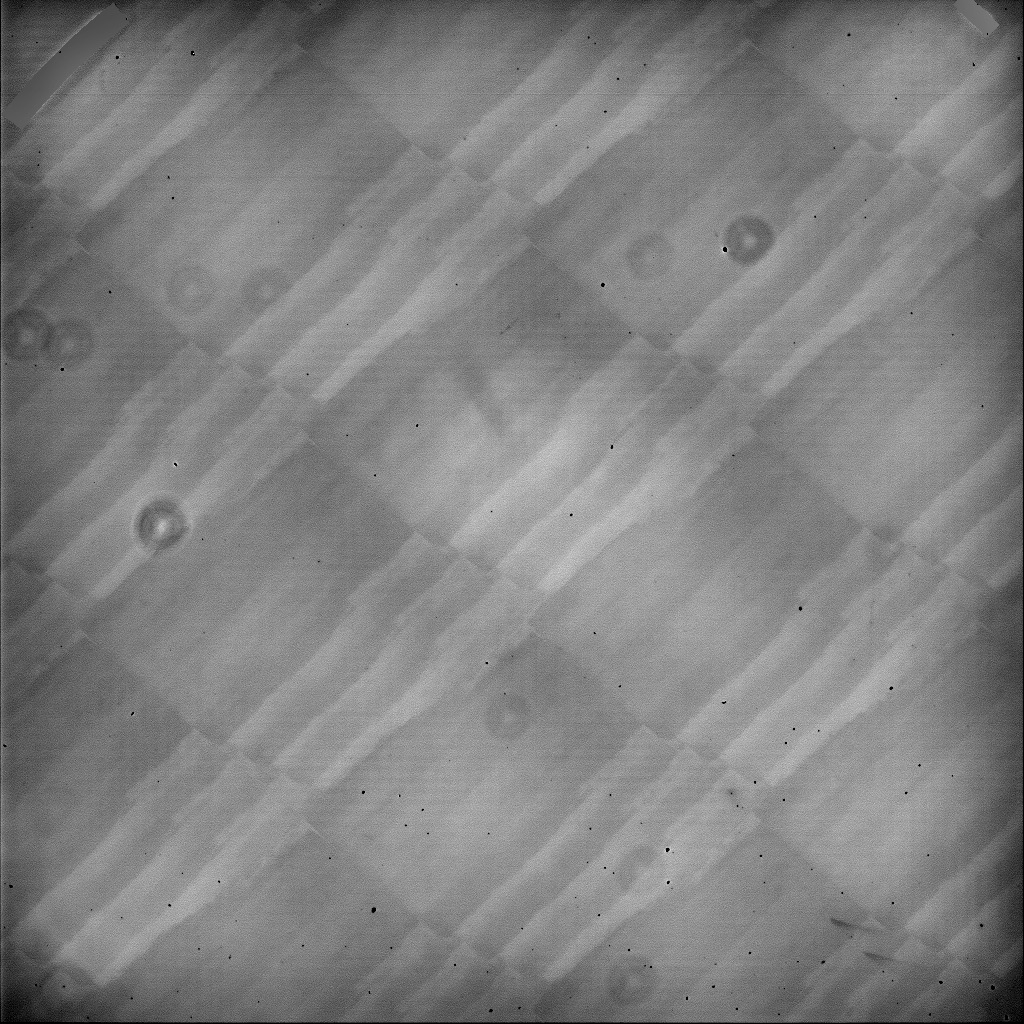}
\caption{The LORRI \onebyone flat-field image used in the calibration pipeline is displayed
using a linear intensity stretch from 0.95 to 1.05.
This flat was produced by co-adding 100 panchromatic images taken during the ground calibration \citep{morgan:2005}.
The diagonal-shaped structures are intrinsic to the CCD.
The small black spots are paint flakes that rained down on the CCD
during the LORRI vibration test.
The larger donut-shaped features are probably out-of-focus flakes (e.g., flakes on one of the lens surfaces).
The flakes themselves were likely produced when the last lens in the lens assembly was replaced
because of a chipped edge, during the latter stage of the telescope assembly.
Most of the flakes were blown off the CCD after the vibration test, but many remained essentially stuck
to the CCD and haven't moved over the entire duration of the \nh mission (as documented from
images of the internal calibration lamp).
The dark streaks in the lower right quadrant were apparently produced when a camel hair brush was moved
across the CCD (in an attempt to remove the flakes).
There are two diagonal-shaped artifacts, one in the upper left (the larger one) and one in the upper right,
which are the residuals from our attempt to remove the stray light seen in those regions.
}
\label{fig:ff}
\end{figure}

No suitable flat field astronomical target was available after launch.
The flat-field reference file used in the LORRI calibration pipeline was produced by averaging 100 flat-field images taken at room
temperature during ground testing using the xenon arc lamp as the light source, debiasing and desmearing the average image as described
earlier, and normalizing the intensities in the active region to a median value of 1 \mbox{(Figure \ref{fig:ff}).} 
If S (units are DN) is an image of a target that has already been desmeared and debiased, 
and if FF is the reference flat-field image, then the flat-fielded (i.e., photometrically-corrected) 
target image (C; units are DN) is given by \mbox{C $=$ S/FF.}

\begin{figure}[htb!]
\includegraphics[keepaspectratio,width=\linewidth]{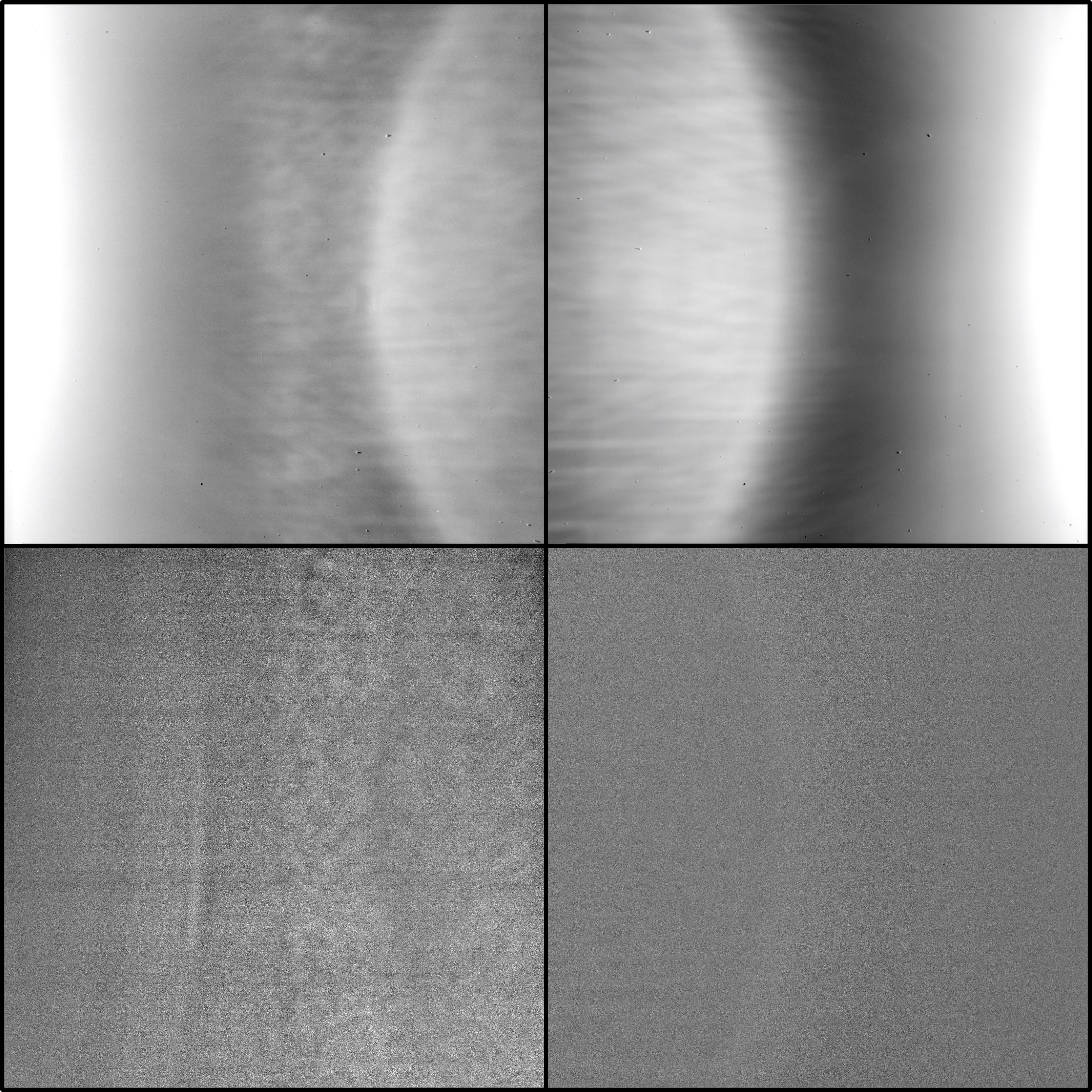}
\caption{Two ``cal lamps'' provide illumination of the LORRI CCD during in-flight functional tests.
The upper left frame shows a \onebyone CCD image during illumination from lamp~\#1,
and the upper right frame shows a \onebyone image during illumination from lamp~\#2,
both displayed using a hyperbolic sine intensity stretch to better show the dynamic range in the images.
Both images were taken during ACO-1 on 2007 October~29, and both show the highly non-uniform
light distribution produced by these lamps.
The lamps illuminate multiple particles that rained down on the CCD during a vibration test.
The oblique illumination across the CCD produces shadows for some of the images of these particles.
Although highly non-uniform spatially, the illumination from the lamps has been stable over time.
The bottom frames show the ratios of the ACO-1 lamp images to the ACO-8 lamp images taken on 2014 July~5,
for lamp~\#1 (lower left frame) and lamp~\#2 (lower right frame).
Both ratio images are displayed using a linear intensity
stretch ranging from 0.9 to 1.1 (i.e., a total range of $\pm$10\%).
}
\label{fig:lamps}
\end{figure}

The LORRI flat-field has been monitored throughout the \nh mission by taking exposures with onboard ``cal lamps'', 
which refer to two small tungsten filament lamps mounted on opposite sides of the CCD,  in close proximity to the CCD.
The primary function of these onboard lamps is to provide illumination of  the CCD during ``functional'' tests, which verify basic
operation of the CCD and its electronics, but \emph{not} the OTA.
The illumination provided by the lamps is highly non-uniform spatially (see the top row of Figure~\ref{fig:lamps}) but is highly stable over time.
The temporal stability of the lamps has been monitored annually throughout the mission.
Ratios of lamp images taken at different times provide a sensitive measure of the stability 
of the lamps (see the bottom row of Figure~\ref{fig:lamps}).
The histogram for the lamp~\#2 ratio image in Figure~\ref{fig:lamps} is a bit tighter than the histogram for the lamp~\#1 ratio image, but both lamps
show remarkable temporal stability with the average, median, and most common ratio values within 0.1\% of unity for both lamps.
Furthermore, these lamp images demonstrate that the particles on the CCD have not moved over the entire
duration of the mission (i.e., their locations are fixed in CCD coordinates), which means the flat-field correction
employed by the calibration pipeline accurately removes any artifacts they might create during observations.

\subsection{Conversion to Scientific Units} \label{subsec:convert}

The software pipeline that performs the calibration steps defined above does not perform the conversion from 
DN to physical units because that conversion requires knowledge of the spectral energy distribution (SED) of the target.
(The SED is related to the ``color'' of the target in standard astronomical usage.)
Instead, various LORRI FITS header keywords
(``photometry'' keywords) are provided that allow users to convert from DN to physical units depending on the spectral
type and spatial distribution (diffuse vs point source) of the target.
Photometry keywords are provided for targets having spectral distributions similar to Pluto, Charon, Pholus, Jupiter, MU69, and
the Sun. The units adopted for the radiance (also called ``intensity'') of diffuse targets are 
\mbox{ergs cm$^{-2}$ s$^{-1}$ \AA$^{-1}$ sr$^{-1}$ .} 
The units adopted for
the irradiance (also called ``flux'') of point (i.e., unresolved) targets are \mbox{ergs cm$^{-2}$ s$^{-1}$ \AA$^{-1}$.} 
The latest (i.e., current) values of the photometry keywords
are provided in the header of the calibrated (called ``Level 2'') FITS file for the image being analyzed.
The photometry keywords derived from the in-flight calibration campaign conducted in July~2016,
using the star HD~37962 (with a solar-type SED) as the absolute calibration standard, is provided in Table~\ref{tab:photometry}.
All of these keywords enable conversion of raw signals in engineering units to absolute signals
at the so-called ``pivot'' wavelength ($\lambda_{pivot}$), which is one way of characterizing the ``effective'' wavelength
for a broadband optical instrument. 
The pivot wavelength is defined as:
\begin{equation}
\lambda_{pivot} = \sqrt{ \frac{\int \mathrm{QE} \ast \lambda \; d\lambda}{\int  \mathrm{QE}/\lambda \; d\lambda} }
\end{equation}
where ``QE'' is the total system quantum efficiency (see the next section).
The pivot wavelength for LORRI is calculated to be 6076~\AA.

\begin{deluxetable*}{lcc}
\tablecaption{LORRI photometry keywords\label{tab:photometry}}
\tablewidth{0pt}
\tablehead{
\colhead{Keyword} & \colhead{Value (\onebyone)} & \colhead{Value (\fourbyfour)} \\
}
\startdata
RSOLAR & $2.349 \times 10^{5}$ & $4.092 \times 10^{6}$ \\
RPLUTO & $2.270 \times 10^{5}$ & $3.955 \times 10^{6}$ \\
RCHARON & $2.318 \times 10^{5}$ & $4.039 \times 10^{6}$ \\
RJUPITER & $2.069 \times 10^{5}$ & $3.605 \times 10^{6}$ \\
RMU69 & $2.499 \times 10^{5}$ & $4.354 \times 10^{6}$ \\
RPHOLUS & $2.724 \times 10^{5}$ & $4.746 \times 10^{6}$ \\
 & & \\
PSOLAR & $9.533 \times 10^{15}$ & $1.038 \times 10^{16}$ \\
PPLUTO & $9.214 \times 10^{15}$ & $1.003 \times 10^{16}$ \\
PCHARON & $9.410 \times 10^{15}$ & $1.025 \times 10^{16}$ \\
PJUPITER & $8.397 \times 10^{15}$ & $9.144 \times 10^{15}$ \\
PMU69 & $1.104 \times 10^{16}$ & $1.105 \times 10^{16}$ \\
PPHOLUS & $1.106 \times 10^{16}$ & $1.204 \times 10^{16}$ \\
\enddata
\tablecomments{The keywords starting with ``R'' are diffuse target sensitivity keywords and their
values have units of \mbox{(DN s$^{-1}$ pixel$^{-1}$) / (ergs cm$^{-2}$ s$^{-1}$ \AA$^{-1}$ sr$^{-1}$).}
The keywords starting with ``P'' are point target sensitivity keywords and their
values have units of \mbox{(DN s$^{-1}$ ) / (ergs cm$^{-2}$ s$^{-1}$ \AA$^{-1}$).}
For point targets the signal refers to values integrated over the entire instrumental PSF.
}
\end{deluxetable*}

For convenience to users, we provide a prescription for converting LORRI signal rates to standard $V$ magnitudes
in the Johnson photometric system:
\begin{equation}
\mathrm{
V = -2.5 \log (S/t_{exp}) +ZPT + CC - AC
}
\label{eqn:vmag}
\end{equation}
where V is the magnitude in the standard Johnson $V$ band (i.e., specifies the target's flux at 5500~\AA),
S is the measured signal in the selected photometric aperture (DN), t$_{\mathrm{exp}}$ is the exposure time (s),
$\mathrm{ZPT}$ is the photometric zero point (18.78 for \onebyone and 18.88 for \fourbyfour),
$\mathrm{CC}$ is a color correction term (as specified in Table~\ref{tab:vphot}), 
and $\mathrm{AC}$ is an aperture correction term to convert from the flux collected in a specified synthetic aperture 
to the total flux integrated over the LORRI PSF.
The $\mathrm{CC}$ terms listed in Table~\ref{tab:vphot} were calculated using SEDs
for the listed spectral types.
For the specified stellar types, we used SEDs downloaded from the exposure time calculators at the Space Telescope Science Institute.
For the other listed spectral types, we used SEDs adopted by the \nh team, which are based on published spectra. 
For typical LORRI observations of point sources, the SNR is optimized by integrating over a circular aperture
with a radius of 5 pixels (\onebyone format) or 3 pixels (\fourbyfour format), in which case
$\mathrm{AC}$ is either 0.10 or 0.05, respectively.

\begin{deluxetable*}{ll}
\tablecaption{LORRI color corrections \label{tab:vphot}}
\tablewidth{0pt}
\tablehead{
\colhead{Spectral Type} & \colhead{$\mathrm{CC}$}
}
\startdata
O, B, A stars	& $-0.060$ \\
F, G stars		& $+0.000$ \\
K stars		& $+0.400$ \\
M stars		& $+0.600$ \\
Pluto			& $-0.037$ \\
Charon		& $-0.014$ \\
Jupiter		& $-0.138$ \\
Pholus		& $+0.161$ \\
MU69		& $+0.067$ \\
\enddata
\tablecomments{For targets of the specified spectral type, $\mathrm{CC}$ provides the color correction term in the formula
used for converting LORRI count rates \mbox{(DN s$^{-1}$)} to Johnson $V$ magnitude.
}
\end{deluxetable*}

We provide here two examples showing how to convert from engineering units to physical units: one for a diffuse target and
one for a point (i.e., unresolved) target.
Consider a diffuse target whose spectrum is similar to that of Pluto's globally averaged SED.
In this case, the RPLUTO photometry keyword in the header of the Level~2 (i.e., calibrated) file should be used
to convert from the observed count rate in a pixel to a radiance value at LORRI's pivot wavelength:
\begin{equation}
\mathrm{
I = S / t_{exp} / RPLUTO \  (diffuse \ target)
}
\label{eqn:dexample}
\end{equation}
\noindent where:\\

\noindent I is the diffuse target radiance (ergs cm$^{-2}$ s$^{-1}$ \AA$^{-1}$ sr$^{-1}$) at $\mathrm{\lambda_{pivot}}$\\
S is the measured signal in a pixel (DN) \\
$\mathrm{t_{exp}}$ is the exposure time (s) \\
RPLUTO is the LORRI diffuse photometry keyword for targets with Pluto-like SEDs \\

\noindent Since the solar flux ($\mathrm{F_{\sun}}$) at a heliocentric distance of 1~AU at the LORRI pivot wavelength is 
\mbox{176 ergs cm$^{-2}$ s$^{-1}$ \AA$^{-1}$,} the value for the
radiance can be converted to I/F (where $\mathrm{\pi F = F_{\sun}}$), which is a standard photometric quantity used in planetary science, using:
\begin{eqnarray}
\mathrm{
I/F =  \pi Ir^{2} /  F_{\sun} }\\
\mathrm{
\rightarrow I/F = (S / t_{exp} / RPLUTO) \ast \pi  r^{2} /  F_{\sun}
}
\label{eqn:iof}
\end{eqnarray}
where ``r'' is the target's heliocentric distance in AU.

For unresolved targets (e.g., planetary targets observed at large ranges), 
the absolutely calibrated flux (also called the ÒirradianceÓ) at the LORRI pivot wavelength
can be determined using the point source photometry keywords. 
For a target with an SED similar to that of MU69, 
the observed count rate integrated over the LORRI PSF can be related 
to the flux (not to be confused with ``F'' in ``I/F'') at the LORRI pivot wavelength by:
\begin{equation}
\mathrm{
F = S_{total} / t_{exp} / PMU69 \  (point \ target)
}
\label{eqn:pexample}
\end{equation}
\noindent where:\\

\noindent F is the point target flux, or irradiance (ergs cm$^{-2}$ s$^{-1}$ \AA$^{-1}$) \\
$\mathrm{S_{total}}$ is the total signal from the target integrated over the PSF (DN) \\
$\mathrm{t_{exp}}$ is the exposure time (s) \\
PMU69 is the LORRI point source photometry keyword for targets with MU69-like SEDs \\

When observing stars, it is more common to convert the absolute flux to a magnitude 
in a standard photometric system.
For an A-type star observed by LORRI, the $V$ magnitude is given by:
\begin{equation}
\mathrm{
V_{star} = -2.5 \log (S_{total}/t_{exp}) + ZPT - 0.060
}
\label{eqn:vstar}
\end{equation}
where $\mathrm{V_{star}}$ is the star's magnitude in the standard Johnson $V$ band,
$\mathrm{S_{total}}$ is the total signal integrated over the LORRI PSF (DN), 
$\mathrm{t_{exp}}$ is the exposure time (s),
$\mathrm{ZPT}$ is the photometric zero point (18.78 for \onebyone and 18.88 for \fourbyfour),
and the color correction term is $-0.060$.

\section{In-Flight Calibration Measurements and Results} \label{sec:obs}

In-flight LORRI calibration measurements started shortly after the launch of the \nh spacecraft on
\mbox{2006 January 19} and have continued at regular intervals throughout the mission.
The relevant calibration measurements conducted during the mission are listed
in Table~\ref{tab:calobs}.
The results from these observations are summarized in the following sub-sections.

\begin{deluxetable*}{llcll}
\tablecaption{LORRI in-flight calibration observations\label{tab:calobs}}
\tablewidth{0pt}
\tablehead{
\colhead{Cal ID} & \colhead{Date} & \colhead{SAP ID} & \colhead{Target} & \colhead{Objective} 
}
\colnumbers
\startdata
ACO-0 & 2006 Apr 23,24 & 025 & N/A & Bias images, CR monitoring (\fourbyfour) \\
 & 2006 May 23 & 025 & N/A & Bias images, CR monitoring (\fourbyfour) \\
 & 2006 Jul 30 & 006 & N/A & Bias images, CR monitoring (\onebyone) \\
 & 2006 Aug 29 & 007 & M7 & After door opened for first time \\
 & 2006 Aug 31 & 010 & M7 & Optical performance, Linearity, Pointing drift \\
 & 2006 Aug 31 & 018 & M7 & Coalignment measurements with MVIC \\
 & 2006 Sep 04 & 027 & Jupiter & Test of short exposure times ($\leq$ 10 ms) \\
 & 2006 Sep 29 & 013 & N/A & Solar scattered light (3-axis) \\
 & 2006 Sep 24 & 020 & M7 & Mosaic and geometric distortion \\
 & 2007 Jan 10 & 023 & Callirrhoe & First test of RCM ($\mathrm{t_{exp} = 5, 10\ s}$) \\
ACO-1 & 2007 Sep 29 & 013 & N/A & Solar scattered light (3-axis) \\
 & 2007 Oct 29 & 050 & N/A & Functional test \\
ACO-2 & 2008 Oct 13 & 047 & N/A & Solar scattered light (spinning) \\
 & 2008 Oct 15 & 043 & N/A & Functional test \\
ACO-3 & 2009 Jul 21 & 050 & N/A & Functional test \\
ACO-4 & 2010 Jun 24 & 047 & N/A & Solar scattered light (spinning)\\
 & 2010 Jun 25 & 043 & M7 & Optical performance, Functional \\
ACO-5 & 2011 May 23 & 050 & N/A & Functional test \\
ACO-6 & 2012 May 23 & 050 & N/A & Functional test \\
 & 2012 Jun 01 & 055 & M7 & Optical performance \\
ACO-7 & 2013 Jun 22 &  050 & N/A & Functional test \\
 & 2013 Jul 02 & 080 & M7 & Optical performance \\
 & 2013 Jul 03 & 081 & NGC 3532 & Optical performance \\
ACO-8 & 2014 Jul 05 & 050 & N/A & Functional test \\
 & 2014 Jul 22 & 082 & NGC 3532 & Optical performance \\
Cal Campaign & 2016 Jul 03 & 050 & N/A & Functional test \\
 & 2016 Jul 11 & 081 & NGC 3532 & Optical performance \\
 & 2016 Jul 14 & 102 & HD 37962 & Absolute calibration \\
 & 2016 Jul 16 & 103 & HD 205905 & Absolute calibration \\
 & 2017 Sep 18 & 081a & NGC 3532 & Verify performance after 100 days without decontamination\\
 & 2017 Dec 05 & 081c & NGC 3532 & Verify performance after 180 days without decontamination\\
\enddata
\tablecomments{``ACO'' stands for ``Annual Check Out''.
``SAP'' stands for ``Science Activity Plan''.
``CR'' stands for ``Cosmic Ray''.
``RCM'' stands for ``Relative Control Mode''.
``Functional'' tests include bias images (0~ms exposure times), lamp exposures,
and autoexposure images.
All listed dates are in UTC in the spacecraft frame.
}
\end{deluxetable*}

\subsection{Optical Performance} \label{subsec:optical}

LORRI's optical performance has been monitored throughout the mission by observing
star clusters.
Measurements of the two-dimensional spatial distributions of individual stars
enable characterization of the LORRI PSF across the entire FOV.
Photometry of the individual stars is performed to monitor LORRI's sensitivity,
both across the FOV and as a function of time.
By using clusters whose stars are catalogued in one or more astrometric surveys, LORRI's geometrical distortion can
also be measured and monitored as a function of time.  

\begin{figure}[htb!]
\includegraphics[keepaspectratio,width=\linewidth]{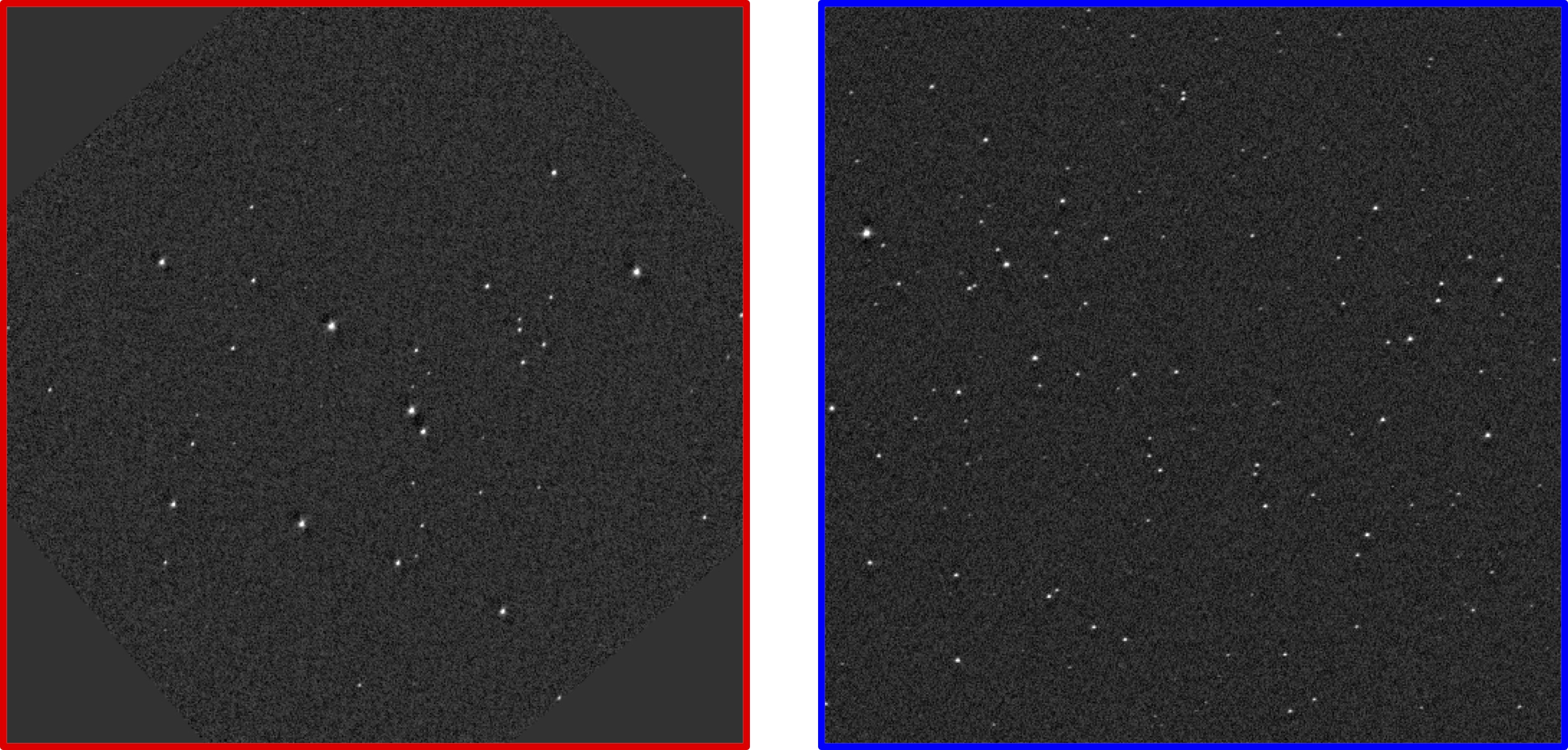}
\caption{LORRI \onebyone images of M7 (left) and NGC~3532 (right) are displayed using the same
hyperbolic sine ($\sinh$) intensity stretch ranging from $-$0.2 to 100~DN.
Both images are 100~ms exposures and are displayed with celestial north up
and east left.
Although M7 has brighter stars, NGC~3532 has a higher spatial density of stars,
which provides better spatial coverage over the full LORRI field-of-view.
}
\label{fig:m7ngc}
\end{figure}

The galactic open cluster Messier~7 (M7; alternate names are NGC~6475 and the Ptolemy cluster)
was used for LORRI's optical performance calibrations from 2006 through 2013.
In 2013, we switched to the galactic open cluster NGC~3532 (alternate names are the Wishing Well cluster, 
the Pincushion cluster, and the Football cluster), primarily because NGC~3532 has a higher
density of stars (Figure~\ref{fig:m7ngc}), which allows better areal coverage over the full CCD and improved mapping of LORRI's geometrical
distortion.
We observed \emph{both} clusters during ACO-7 in 2013 so that later observations of NGC~3532 (i.e., those taken after 2013) could be compared
to earlier observations of M7 (i.e., those taken between 2006 and 2013), thereby enabling systematic
monitoring of LORRI's performance over the entire mission.

\begin{figure}[p!]
\includegraphics[keepaspectratio,width=\linewidth]{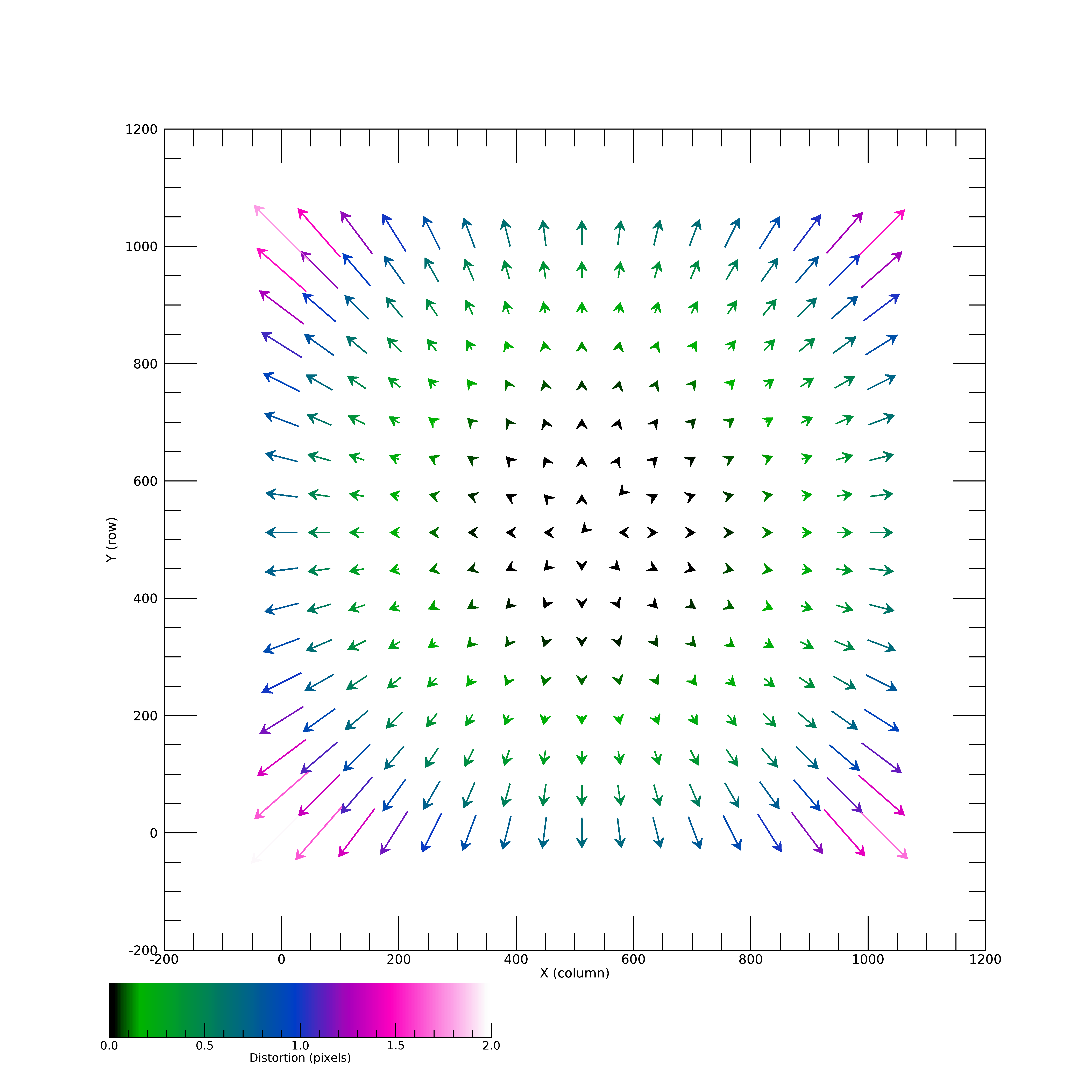}
\caption{LORRI \onebyone geometrical distortion map derived from the analysis
of an image of NGC~3532.
At each plotted location on the CCD (i.e., CCD pixel coordinates), the arrow points in the direction of the distortion
and its size  is proportional to the magnitude of the distortion.
The colors of the arrows encode the magnitude of the distortion in \onebyone pixels,
as given in the colorbar.
LORRI has a small amount of pin-cushion distortion, with the largest distortion
occurring in the four corners, where the magnitude grows to ~1.5~pixels. 
}
\label{fig:distortion}
\end{figure}

The optical design of LORRI has a small amount of pin-cushion distortion, which is clearly detected in 
measurements of the star clusters (Figure \ref{fig:distortion}).
However, LORRI achieved its design goal of keeping the geometrical distortion
$\leq$0.3\% over the entire FOV.
Spot checks of the geometrical distortion at different times in the mission have not shown any significant differences 
(i.e., the various distortion parameters don't change by more than their measurement uncertainties, at the 2$\sigma$ level).
The LORRI geometrical distortion is described in detail in the SPICE instrument kernel for LORRI, which is archived
at the Small Bodies Node (SBN) of the Planetary Data System (PDS).
The geometrical distortion coefficients described within the Simple Image Polynomial (SIP) framework, which is commonly
used in astronomy, are captured in the LORRI FITS header keywords.
We note also that the archived LORRI FITS files employ World Coordinate System (WCS) keywords 
to enable accurate transformations between native LORRI [x,y] pixel locations and standard astronomical coordinates (e.g., [RA,DEC]).

\begin{figure}[p!]
\includegraphics[keepaspectratio,width=\linewidth]{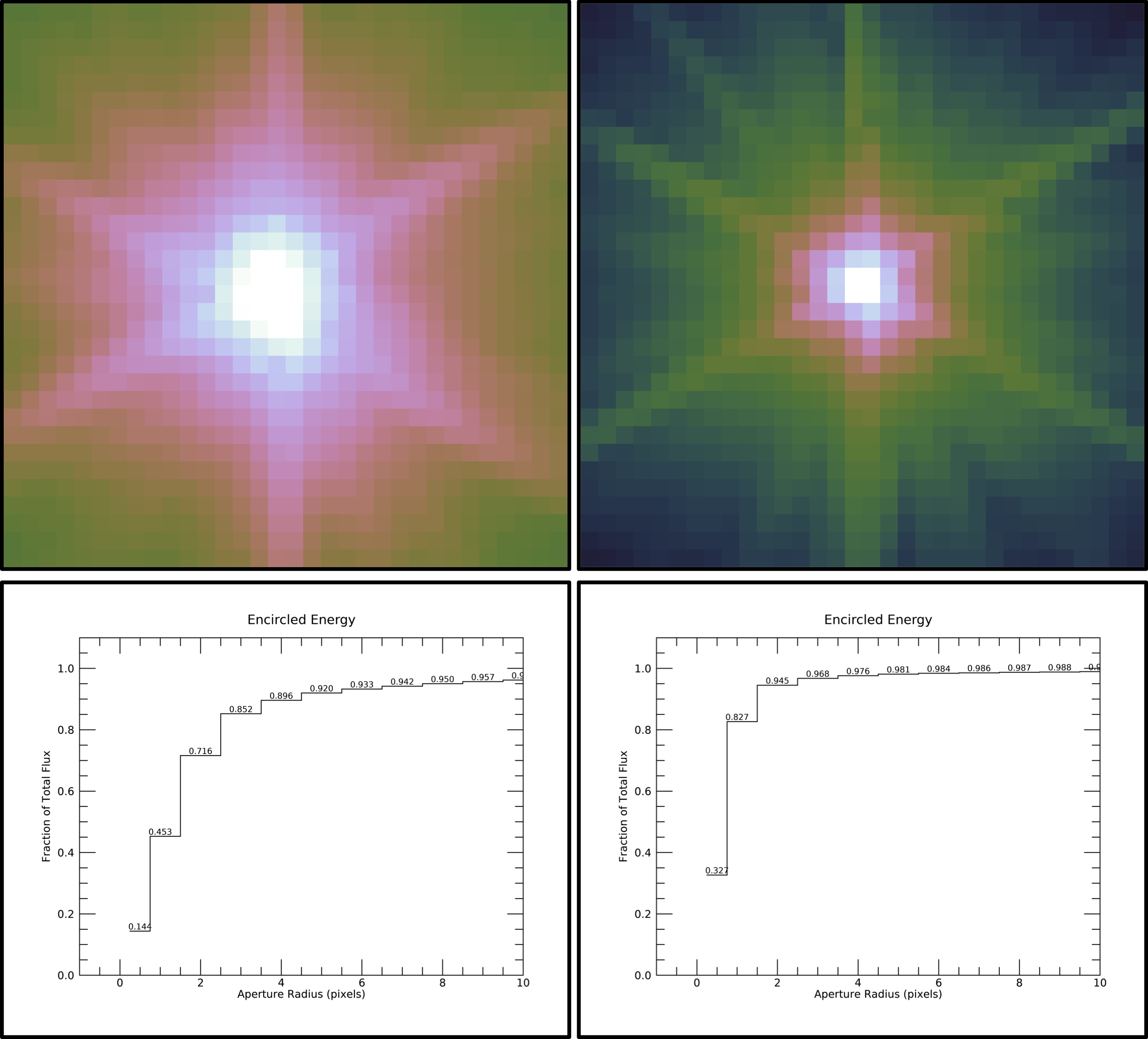}
\caption{LORRI composite PSFs, valid for locations near the center of the CCD, are displayed for 
both \onebyone (upper left) and \fourbyfour (upper right) CCD formats.
In both cases, \mbox{32 $\times$ 32} pixel regions are displayed using a hyperbolic sine ($\sinh$) intensity stretch to show more
clearly the full dynamic range of the image.
Diffraction spikes from the three legs of the OTA spider are clearly evident.
Encircled energy (EE) plots are displayed below each image; each data point is labeled with the fraction of light within the plotted radius. 
In \onebyone format, the peak pixel contains $\sim$14\% of the total intensity. 
In \fourbyfour format, the peak pixel contains $\sim$32\% of the total intensity. 
}
\label{fig:psf}
\end{figure}

High SNR PSFs for both \onebyone and \fourbyfour LORRI images are displayed in Figure~\ref{fig:psf}.
These were created by combining multiple images of single stars located near the center of the LORRI FOV,
including images taken with different exposure times to expand the dynamic range of the composite
(e.g., for \onebyone images, using 100~ms exposures to sample the PSF cores and 
using 400~ms exposures to sample the PSF wings).
Three diffraction spikes produced by the three struts of the OTA are clearly visible.
The PSFs are not completely symmetrical; there is excess flux extending from the peak pixel toward the lower right, which
is presumably produced by a slight misalignment of the OTA.
This asymmetry in the PSF has been present at the same level throughout the mission, 
indicating it is a stable feature, which enables accurate deconvolution techniques to be applied
when attempting to maximize the spatial resolution of LORRI images.

\begin{figure}[htb!]
\includegraphics[keepaspectratio,width=\linewidth]{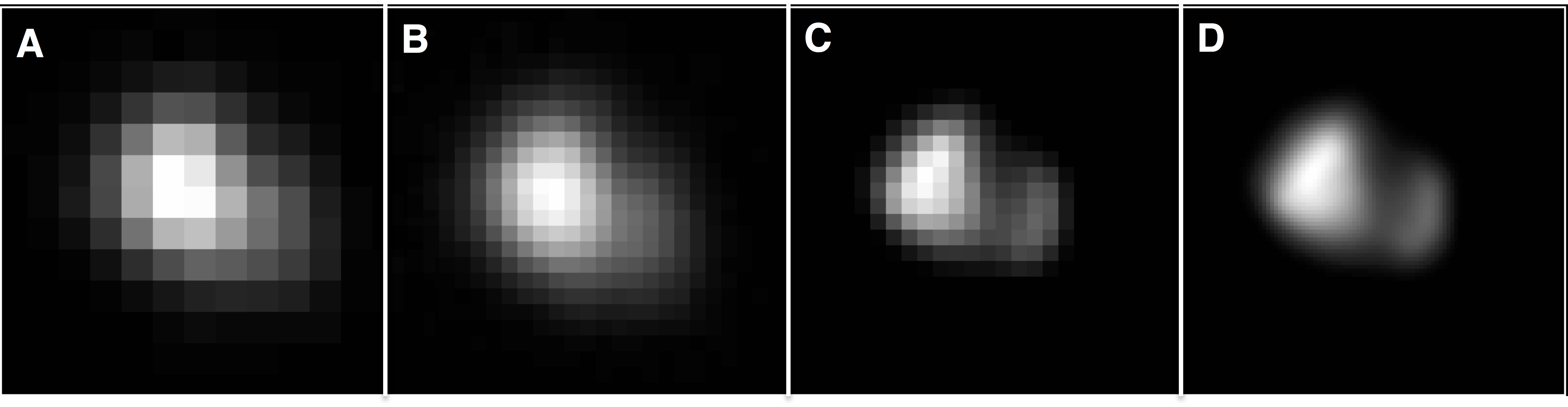}
\caption{Demonstration of the image processing steps performed to produce a deconvolved
LORRI image of Kerberos, one of Pluto's four small satellites.
(A) A single calibrated LORRI image, one of four taken. 
(B) The interlaced and Nyquist-sampled ``superimage'' generated by combining
the four calibrated images -- the pixel scale is twice as fine as the native LORRI scale.
(C) The superimage after applying Lucy-Richardson deconvolution. 
(D) The deconvolved image up-sampled by an additional factor of 4 for a final scale 8 times finer
than the native LORRI scale. 
This latter step removes the pixelated appearance of the previous image for improved clarity, and is 
mathematically justified since the superimage is Nyquist-sampled.
Adapted from \citet{weaver:2016}.
}
\label{fig:kerberos}
\end{figure}

A technique for creating properly sampled (i.e., Nyquist sampled) PSFs is described in \citet{lauer:1999b} and has been
applied to LORRI images when multiple, dithered exposures are available.
A good example is the LORRI imaging of Pluto's satellite Kerberos, when four separate images could be combined
to create a composite with significantly higher resolution than the individual images (Figure~\ref{fig:kerberos}).
Deconvolution can also be used to remove motion smear for LORRI images taken during scanning observations of
the Ralph instrument (e.g., Figure~\ref{fig:chaos}).

Both the shapes and intensities of the stars in the calibration fields are used to determine whether there has been any degradation
in the optical performance over the course of the mission.
For example, contamination (e.g., ice accumulation) anywhere along LORRI's optical path could manifest as a reduction in photometric sensitivity
and/or a broadening of the PSF.
We monitor the PSF shape by fitting 2-dimensional Gaussians to the stellar images each time the calibration fields are observed.
We monitor LORRI's sensitivity by comparing the signals for the same stars measured at multiple epochs.

\begin{figure}[p!]
\includegraphics[keepaspectratio,width=\linewidth]{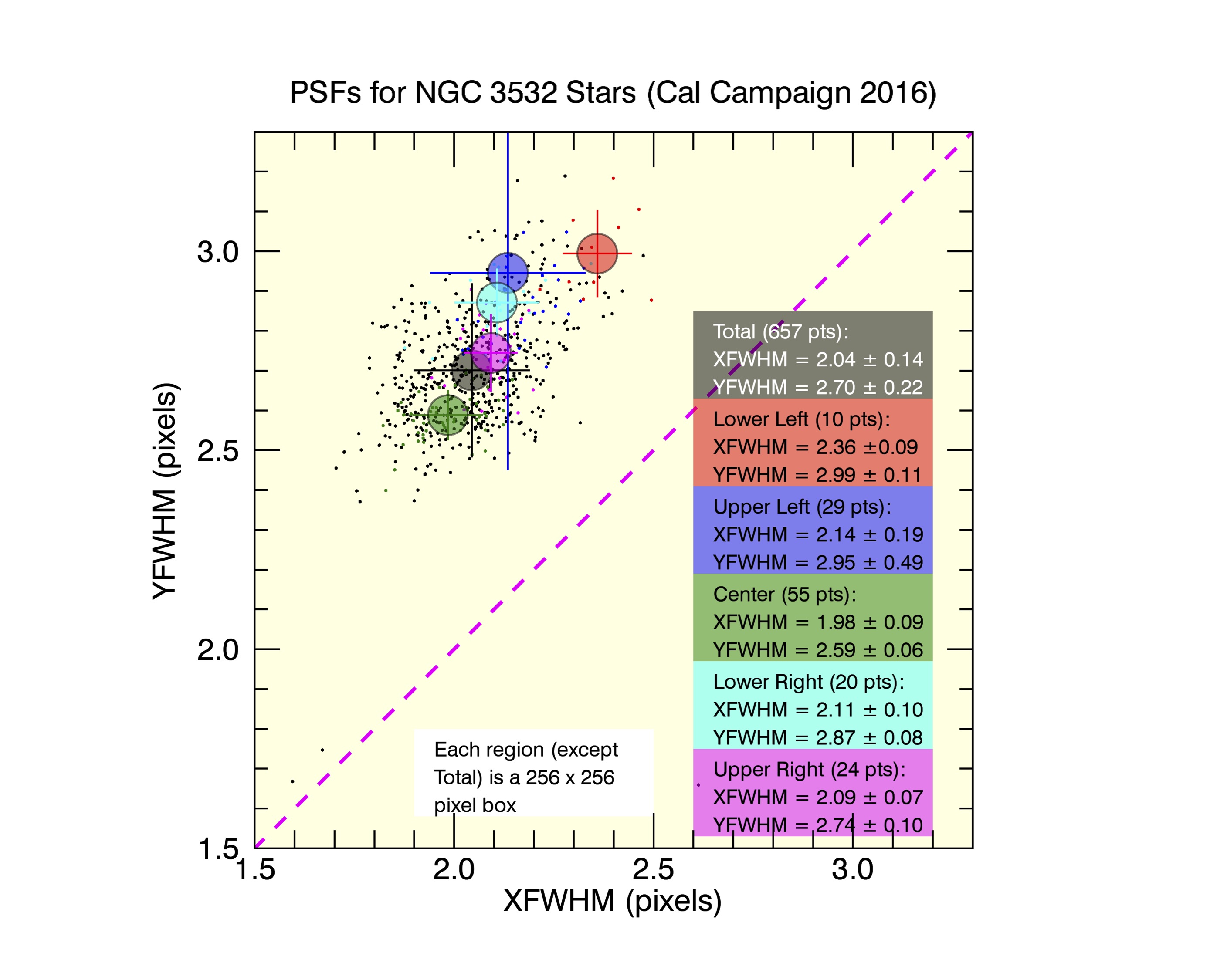}
\caption{Two dimensional Gaussian fits to the spatial brightness distributions
of stars in NGC~3532 observed during the calibration campaign in July~2016
are plotted.
The LORRI boresight was moved around in a \mbox{3 $\times$ 3} raster pattern
centered on the nominal position of NGC~3532, which enabled 
657 separate measurements of the point spread function (PSF) across the CCD.
The $+$x-dimension corresponds to the direction of increasing CCD columns, and
the $+$y-dimension corresponds to the direction of increasing CCD rows.
Sub-groups of the stars have been color coded as a function of their locations on the CCD.
Each sub-group is comprised of stars located in a \mbox{256 $\times$ 256} pixel region 
on the CCD, as indicated in the figure.
The CCD has an optically active region of \mbox{1024 $\times$ 1024} pixels.
The average full width half maximum (FWHM) in each dimension, the standard deviation 
of the FWHMs, and the number of stars in each sub-group, are listed and plotted for each location.
The FWHMs and their standard deviations for all the stars, independent of location,
are also listed and plotted.
The points would lie along the dashed diagonal line if the PSFs had the same widths
in the two directions, but YFWHM is systematically larger than XFWHM due to the
shape of the asymmetric PSF and the better charge transfer efficiency in the x-direction.
See the text for further discussion.}
\label{fig:psf_2016}
\end{figure}

\begin{figure}[htb!]
\includegraphics[keepaspectratio,width=\linewidth]{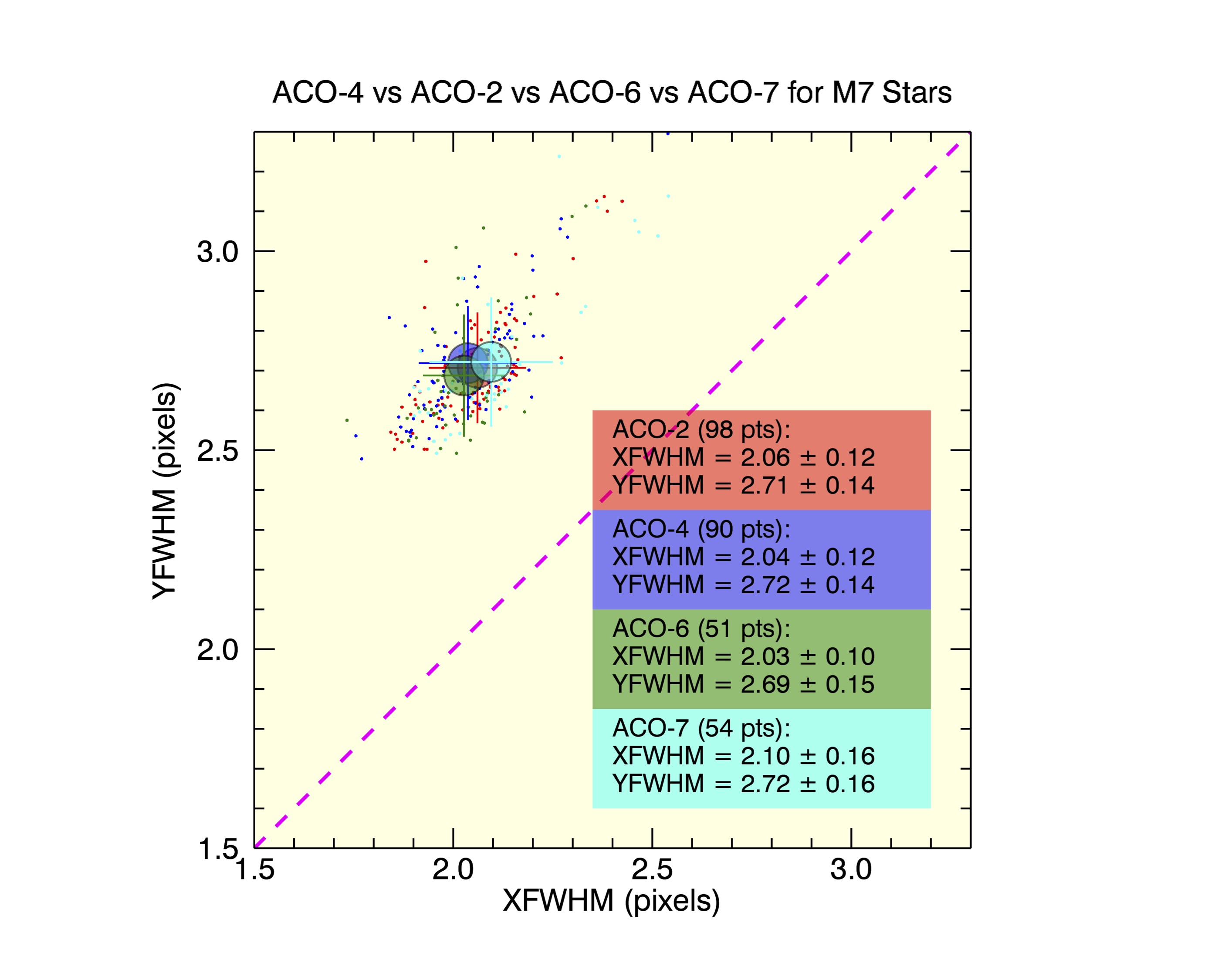}
\caption{Two dimensional Gaussian fits to the spatial brightness distributions
of multiple stars in the open galactic cluster M7 are plotted for four different epochs.
The $+$x-dimension corresponds to the direction of increasing CCD columns, and
the $+$y-dimension corresponds to the direction of increasing CCD rows.
The average full width half maximum (FWHM) in each dimension, and the standard deviation 
of the FWHM, are listed and plotted for each epoch.
ACO is an acronym for ``Annual Check Out''.
See the text for further discussion.}
\label{fig:psf_over_time}
\end{figure}

Figure~\ref{fig:psf_2016} shows the results of the Gaussian fits to the stars measured during the observations of NGC~3532 during
the calibration campaign in July 2016.
By comparing the PSFs from stars falling in five different regions of the CCD, we show how the PSF varies across the LORRI FOV.
The PSF behavior exhibited in this figure is typical of what has been observed throughout the mission.
Figure~\ref{fig:psf_over_time} explicitly compares the LORRI PSFs measured on M7 stars over a 5-year period (2008-2013);
there is no significant variation in the shape of the PSF over this period.

The Gaussian fits applied to the star cluster images employed a $\pm$10~pixel region centered on each star's peak pixel, which includes both the
core and a significant fraction of the wings of the spatial brightness distribution.
Restricting the fit to the core only, which is a better measure of LORRI's spatial resolution,
results in a narrower full width at half maximum (FWHM).
For example, using the same Gaussian fitting routine on the \onebyone PSF displayed in Figure~\ref{fig:psf}
gives \mbox{(XFWHM,YFWHM)$=$(2.06,2.65) pixels} when a $\pm$10~pixel region is fit, and
\mbox{(XFWHM,YFWHM)$=$(1.87,2.47) pixels} when a $\pm$2~pixel region is fit.
Thus, we see that the LORRI PSF is slightly undersampled (relative to Nyquist) in the X (row) direction.

\begin{figure}[p!]
\includegraphics[keepaspectratio,width=\linewidth]{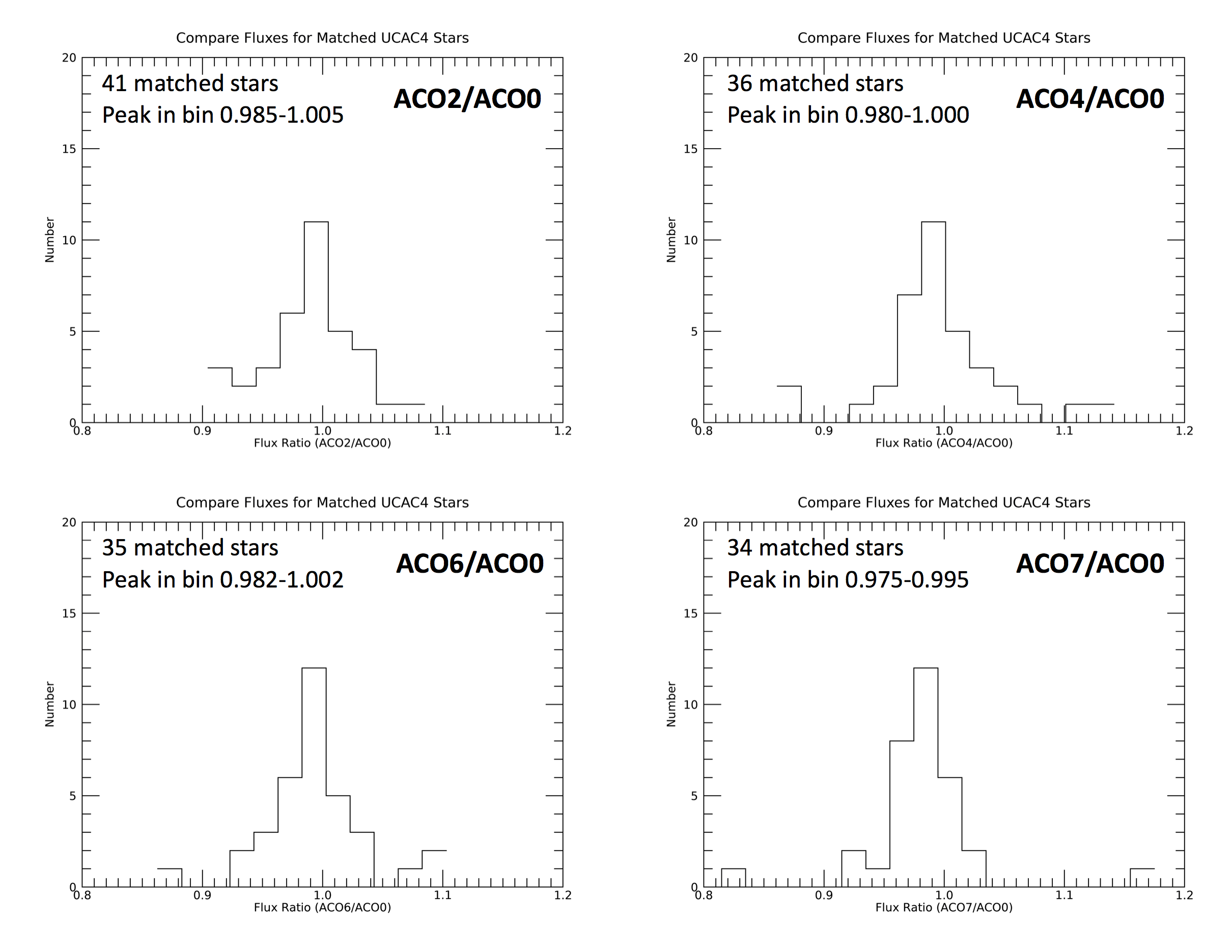}
\caption{LORRI aperture photometry of stars in M7 at four different epochs is compared to that measured
on 2006-August-31 (ACO-0), immediately after the telescope door was opened.
For each epoch, a histogram is plotted showing the ratio of the observed stellar fluxes to the fluxes measured for
those same stars in 2006.
The number of matched stars is displayed on each plot.
The histogram bin widths are 0.02, and the bin location of the peak in the
distribution is listed on each plot.
ACO is an acronym for ``Annual Check Out''.
See the text for further discussion.}
\label{fig:sensA}
\end{figure}

\begin{figure}[htb!]
\includegraphics[keepaspectratio,width=\linewidth]{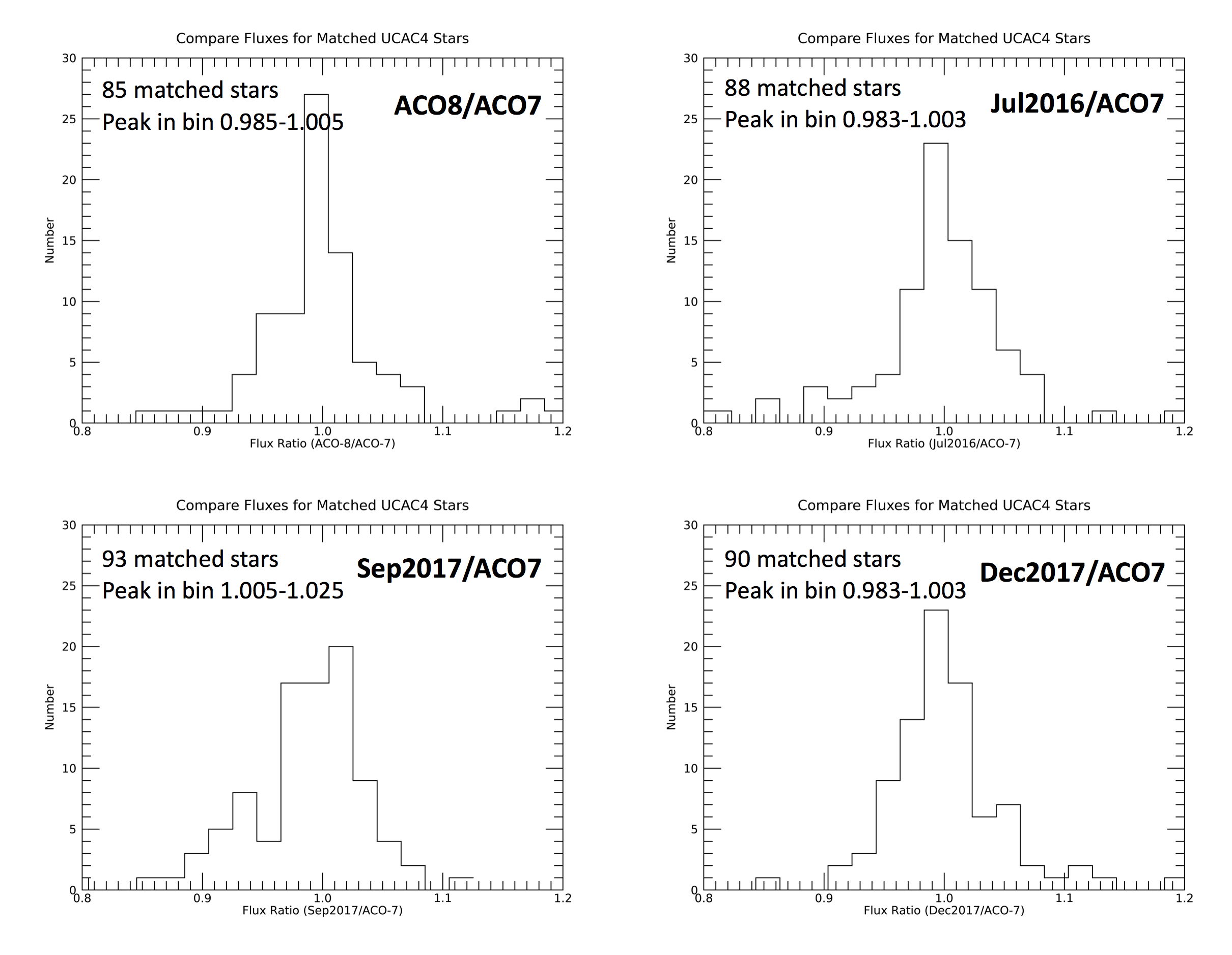}
\caption{LORRI aperture photometry of stars in NGC~3532 at four different epochs is compared to that measured
on 2013-July-03 (ACO-7).
For each epoch, a histogram is plotted showing the ratio of the observed stellar fluxes to the fluxes measured for
those same stars in 2013.
The number of matched stars is displayed on each plot.
The histogram bin widths are 0.02, and the bin location of the peak in the
distribution is listed on each plot.
ACO is an acronym for ``Annual Check Out''.
See the text for further discussion.}
\label{fig:sensB}
\end{figure}

Photometry of M7 stars over a 7-year period (2006-2013) demonstrates that there has been no significant change in
 LORRI's sensitivity during this time (Figure~\ref{fig:sensA}).
 Photometry of NGC~3532 stars over a 4.5-year period (July~2013 to December~2017) also shows no significant
 change in LORRI's sensitivity during this period (Figure~\ref{fig:sensB}).
 Combining all these results, and taking into account that the stars being sampled
 did not necessarily have constant fluxes during the times of the LORRI measurements,
 we conclude that LORRI's sensitivity has remain essentially unchanged (at the level of $\sim$1\%)
 over the entire duration of the mission.

\subsection{Absolute Calibration} \label{subsec:abscal}

Since the SEDs of virtually all solar system targets are produced by scattered sunlight,
we searched for absolute standard stars that are solar analogs to calibrate LORRI.
Fortunately, two solar-type standard stars with absolute fluxes measured to an accuracy
of $\sim$1\% (1$\sigma$) by the \emph{Hubble Space Telescope} (\emph{HST}) have $V$ magnitudes that are suitable
for high SNR LORRI measurements, and they are also visible from the \nh spacecraft
at reasonably large solar elongation angles for the entire mission.

\begin{figure}[htb!]
\includegraphics[keepaspectratio,width=\linewidth]{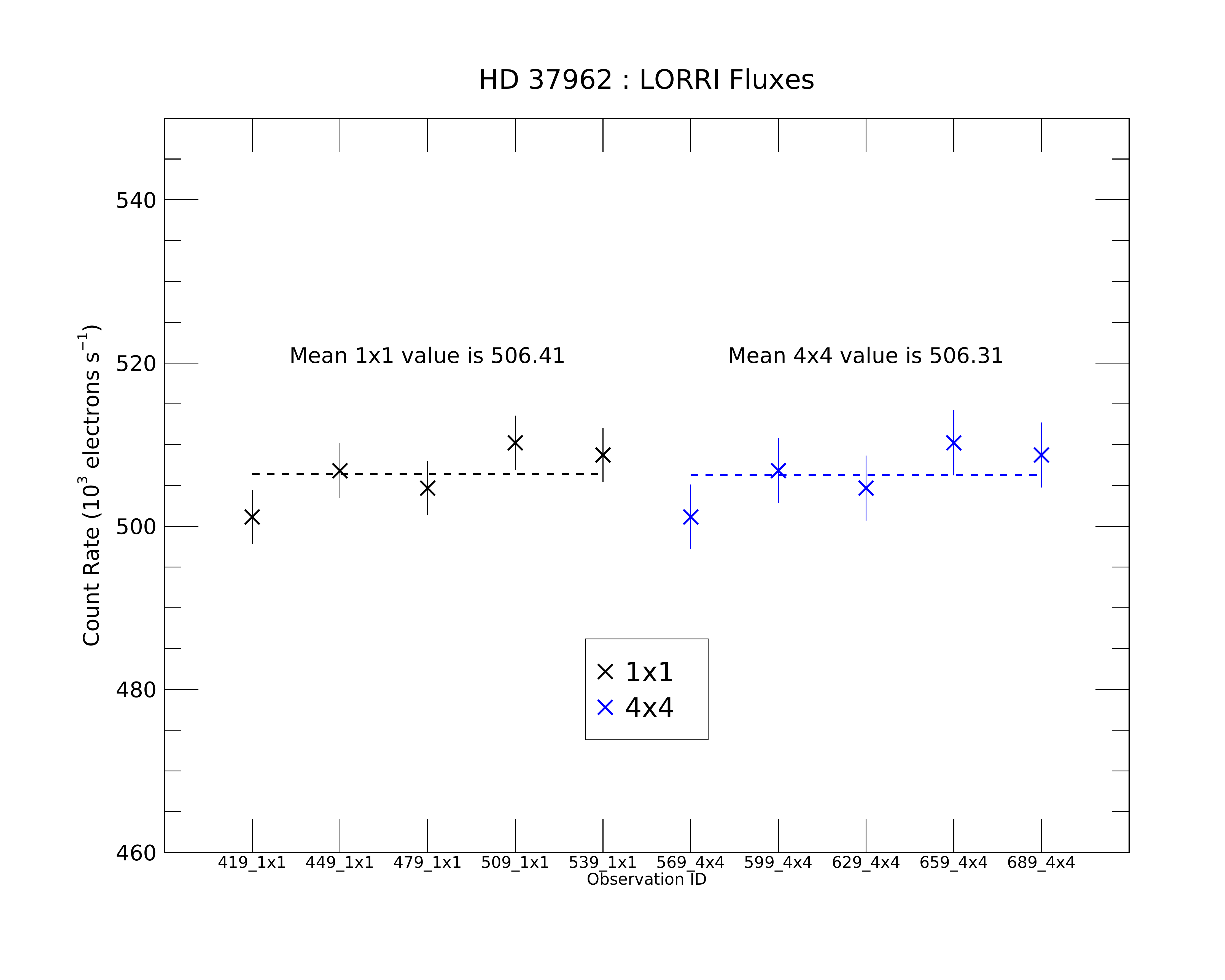}
\caption{The total fluxes from the absolute calibration standard star HD~37962 are plotted
for five different \onebyone images and for five different \fourbyfour images.
For the \onebyone images, the stellar fluxes were derived by integrating the signal in a 5-pixel radius aperture centered on the star and then
subtracting a background level (almost negligible) derived from the average signal in an annulus extending between 10 and 20 pixels from the star.
For the \fourbyfour images, the stellar fluxes were derived by integrating the signal over a 3-pixel radius aperture centered on the star and then
subtracting a background level (almost negligible) derived from the average signal in an annulus extending between 10 and 20 pixels from the star.
These aperture fluxes were then corrected to total fluxes (i.e., integrated over the entire PSF) using the encircled energy curves
presented in Figure~\ref{fig:psf}.
The ``Observation ID'' is a concatenation of the 3 least significant figures in the mission elapsed time (MET) for each observation and
the CCD format used.
The first image in each sequence shows a slightly lower flux than measured in subsequent images, for both \onebyone and \fourbyfour formats,  possibly suggesting a ``start-up'' effect.
However, the effect is small ($\lesssim$1\%) and appears to resolve within $\sim$30~s, which is the spacing between
consecutive images for these observations.
}
\label{fig:calstar_fluxes}
\end{figure}

HD~37962 has $V=7.85$ and $B\!-\!V=0.65$, which is identical to the solar color.
This star was observed by LORRI on 2016 July~14, as part of the special post-Pluto calibration campaign.
The solar elongation angle was 56$\arcdeg$, and the solar scattered light level was negligible in all images.
We obtained 5 different 100~ms exposures in \onebyone format and 5 different 50~ms exposures in \fourbyfour format.
The observed stellar signals had \mbox{SNR $\geq$ 140} in all 10 images.
We used aperture photometry and the gain values discussed previously to calculate the total fluxes 
in \mbox{electrons s$^{-1}$.}
As shown in Figure~\ref{fig:calstar_fluxes}, the measured fluxes for the \onebyone and \fourbyfour images
are in excellent agreement. 

HD~205905 has $V=6.74$ and $B\!-\!V=0.62$, which is slightly bluer than solar color.
This star was observed by LORRI on 2016 July~16, also as part of the special post-Pluto calibration campaign.
We obtained 5 different 100~ms exposures in \onebyone format and 5 different 50~ms exposures in \fourbyfour format.
The observed stellar signals had \mbox{SNR $\geq$ 100} in all 10 images.
The solar elongation angle was 144$\arcdeg$, and the solar scattered light level was negligible in all images.
The photometry results for HD~205905 are fully consistent with the results obtained from HD~37962.

\begin{figure}[htb!]
\includegraphics[keepaspectratio,width=\linewidth]{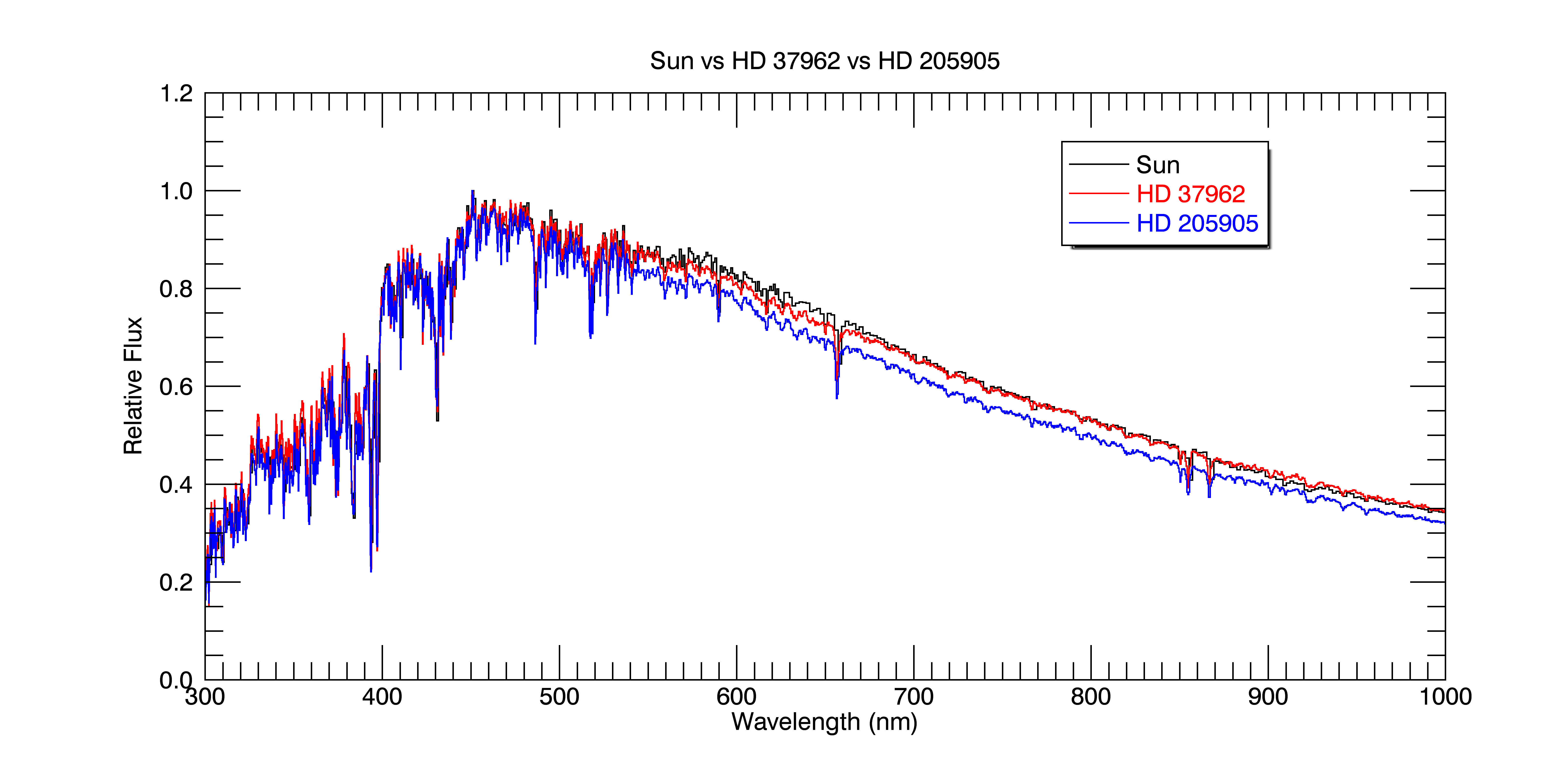}
\caption{The spectral energy distribution (SED) of the two absolute calibration standard stars used
by LORRI (HD~37962 and HD~205905) are compared to the solar spectrum.
All spectra are normalized to their peak values.
The SED of HD~37962 is particularly close to that of the Sun, so this star was selected
as the primary standard star for LORRI absolute sensitivity calibrations.
}
\label{fig:calstars}
\end{figure}

Figure~\ref{fig:calstars} shows a comparison of the stellar and solar spectra.
Given the remarkably solar-like SED for HD~37962, we decided to use that star as the primary
LORRI absolute calibration standard.
Assuming that LORRI's relative responsivity (see the next section) as a function of wavelength was accurately measured
during the ground calibration \citep{morgan:2005}, we adjusted the absolute scale of the responsivity
curve using a single constant multiplicative factor to force the \emph{calculated} signal from HD~37962 
to match the \emph{observed} signal.
The calculated signal is the integral over all wavelengths of the product of the star's SED in absolute units and LORRI's
responsivity curve.
The new LORRI responsivity curve produced in this way is discussed in the next section.
For targets having solar-type SEDs, the absolute accuracy of the values derived from LORRI data should be 
comparable to the accuracy of the \hst measurements (i.e., $\sim$1\%, 1$\sigma$) because the LORRI calibration
is tied to the absolute flux from HD~37962.
However, the accuracy of the LORRI calibration also depends on the accuracy with which the total measured LORRI signal from
HD~37962 is determined.
The peak pixel during these measurements has \mbox{SNR $\geq$ 100}, but we use aperture photometry with a radius
of 5~pixels (for \onebyone images; we used a 3-pixel radius aperture for \fourbyfour images) to measure the signal and 
then use the PSF described in Section~\ref{subsec:optical} to correct to the signal for an infinite aperture (i.e., integrated
over the PSF).
The absolute accuracy of this latter estimation process is probably a few percent.
We note that the \emph{relative} photometry of different solar-type stars is set by the SNR of the individual measurements,
which can be considerably more accurate than the absolute flux measurement.

For targets with SEDs significantly different than solar-type, the LORRI calibration is dependent on the accuracy of
the \emph{shape} of LORRI's responsivity curve, as determined from the ground calibration \citep{morgan:2005}.
As discussed there, we assumed that the wavelength dependence of the responsivity followed the prescriptions
provided by the various vendors contributing to the LORRI hardware.
While these  prescriptions usually involved measurements made on actual LORRI hardware
(e.g., transmittance measurements for the optical components),
we assumed a QE curve for the CCD that followed the ``typical'' response specified by the
manufacturer for the anti-reflection layer ordered for the LORRI CCD.
The shape of the CCD QE curve is thought to be accurate to a few percent in the regions of high QE 
(i.e., for most of the wavelength range covered by LORRI),
but larger variations in QE are possible in the steeply sloped portions of the QE curve.
Given the wide bandpass of LORRI, these latter variations will likely not significantly affect
the measured signals from LORRI's targets. 
For all these reasons, we suggest that the absolute accuracy for LORRI observations of non-solar-type targets
is probably $\sim$10\%, and possibly significantly better than that, approaching the accuracy achieved
for targets with solar-type SEDs.

\subsection{Responsivity Curves} \label{subsec:response}

For a photon counting optical system like LORRI, the signal detected ($\mathrm{S}_\mathrm{e}$ in electrons) in an image pixel 
with exposure time $\mathrm{t}_{\mathrm{exp}}$ (in seconds) can be expressed as:
\begin{equation}
\mathrm{
S_{e} = t_{exp} \ast  A \ast  \Omega \ast  \int_{\lambda} I \ast  QE \ d\lambda \ \  (diffuse \ target)
}
\label{eqn:dsense}
\end{equation}
\noindent and:\\
\begin{equation}
\mathrm{
S_{e} = t_{exp}\ast  A \ast  EE \ast  \int_{\lambda} F \ast QE \ d\lambda \ \ (point \ target)
}
\label{eqn:psense}
\end{equation}
\noindent where:\\

\noindent A is the unobscured aperture area of the OTA (339.8 cm$^{2}$)\\
$\mathrm{\Omega}$ is the IFOV ($2.464 \times 10^{-11}$ sr for \onebyone, $3.942 \times 10^{-10}$ sr for \fourbyfour)\\
I is the diffuse target radiance (photons cm$^{-2}$ s$^{-1}$ \AA$^{-1}$ sr$^{-1}$)\\
F is the point target flux (photons cm$^{-2}$ s$^{-1}$ \AA$^{-1}$)\\
EE is the fraction of the point source flux captured in the peak pixel\\
QE is the total system quantum efficiency\\

The noise (N in electrons) in a LORRI image pixel can be expressed as (for both diffuse and point targets):
\begin{equation}
\mathrm{
N = \sqrt{ S_{e} + SL + FT + (I_{d} \ast  t_{exp}) + RN^{2} }
}
\label{eqn:noise}
\end{equation}
\noindent where:\\

\noindent SL is the signal produced by solar scattered light (e)\\
FT is the signal produced by the CCD frame scrub and transfer process (i.e., smear) (e)\\
$\mathrm{I_{d}}$ is the CCD dark current (e s$^{-1}$ pixel$^{-1}$)\\
RN is the electronics noise (e, including the CCD read noise)\\

\begin{figure}[htb!]
\includegraphics[keepaspectratio,width=\linewidth]{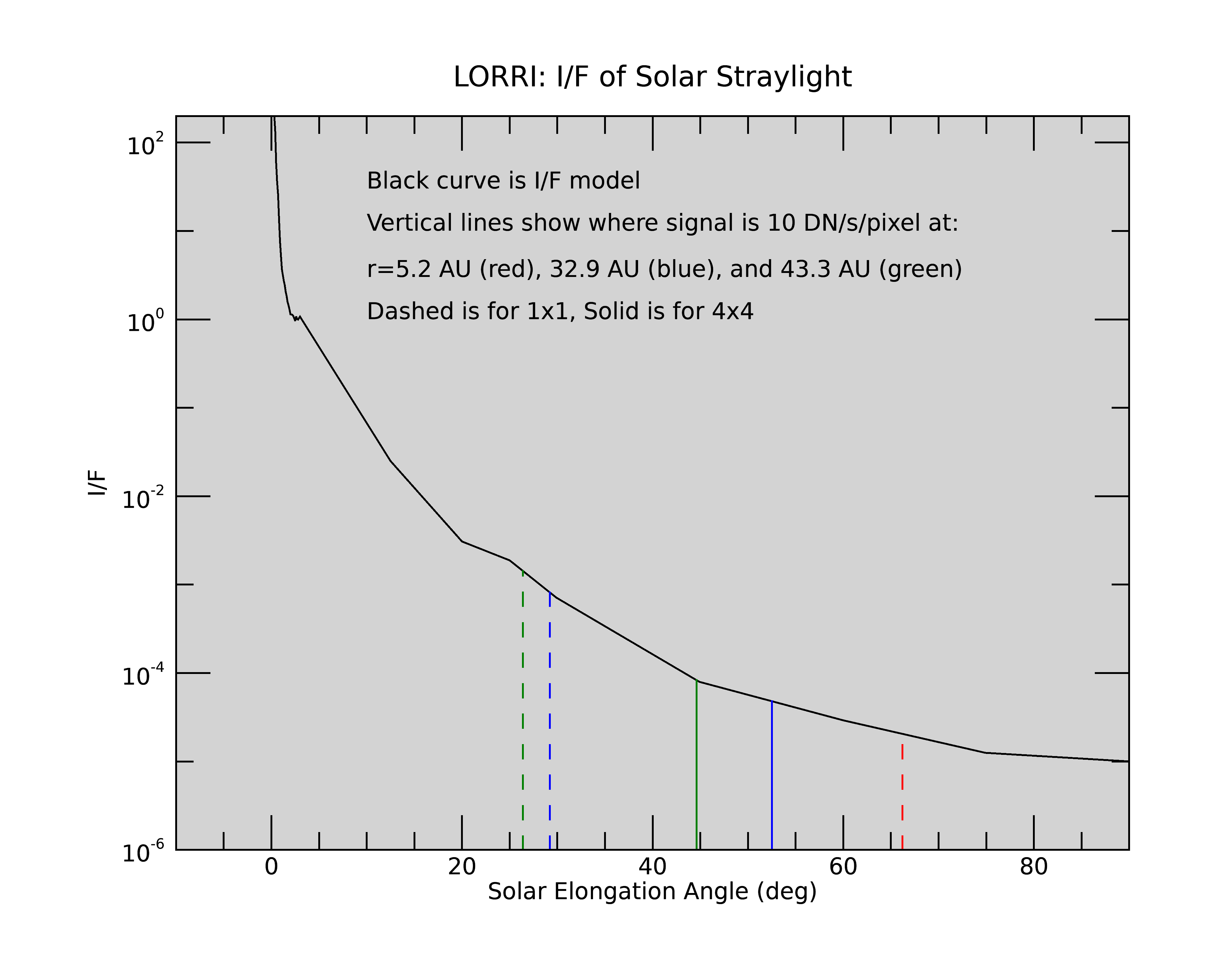}
\caption{A model for the LORRI scattered light level is plotted as a function of solar elongation angle (SEA).
The SEA at which the solar scattered light produces a signal rate of 
\mbox{10 DN s$^{-1}$ pixel$^{-1}$} (i.e., $\sim$10$\times$ larger than the electronics noise) is shown in the figure
for the heliocentric distances of the \nh flybys (i.e., at Jupiter, Pluto, and MU69).
}
\label{fig:scattered_light}
\end{figure}

The LORRI solar scattered light level (SL) has been measured multiple times during the mission (Table~\ref{tab:calobs}).
For solar elongation angles (SEAs) smaller than $\sim$30\arcdeg, the scattered light level varies as a function
of the spacecraft roll angle.
Figure~\ref{fig:scattered_light} shows a model for the I/F of the solar scattered light level as a function of SEA.
The model falls roughly halfway between the smallest and largest observed scattered light values.
At small SEAs, the actual I/F of the solar scattered light could be up to a factor of two times smaller or
larger than the model value.
The model I/F values are independent of the spacecraft's heliocentric distance ($r$), but the CCD signal rate (DN/s)
has an $r^{-2}$ dependence. 
The SEA at which the solar scattered light produces a signal rate of 
\mbox{10 DN s$^{-1}$ pixel$^{-1}$} (i.e., $\sim$10$\times$ larger than the electronics noise) is shown in the figure
for the heliocentric distances of the \nh flybys (i.e., at Jupiter, Pluto, and MU69).

The LORRI team created an exposure time calculator (ETC) that uses the above formalism to
estimate SNRs for planned observations.
The ETC can also be used estimate target signals in absolute units by comparing observed
SNRs to SNRs reported by the ETC.
For point sources with \mbox{SNR $\geq$ 5} in a single pixel, the SNR can typically be improved by a 
factor of $\sim$2 by summing the signal over several pixels (aperture photometry) or by employing 
PSF-fitting photometry. 

The LORRI signal level depends on two key parameters: the total system quantum efficiency (QE) and (in the point source case) 
the amount of energy from a point source concentrated in a single pixel (EE). 
LORRI's EE was discussed in the previous subsection, so we focus on QE here.

For LORRI, the QE is given by:
\begin{equation}
\mathrm{
QE = L_{obscure} \ast T_{optics} \ast QE_{CCD} \\
}
\label{eqn:qe}
\end{equation}

\noindent where:\\

\noindent $\mathrm{L_{obscure}}$ is the loss factor associated with obscuration by the secondary mirror and OTA spider \\
$\mathrm{T_{optics}}$ is the total transmittance of all optical elements \\
$\mathrm{QE_{CCD}}$ is the quantum efficiency of the CCD \\

Although not explicitly stated, all quantities in the equation above are a function of wavelength.
Prior to any system calibration measurements, LORRI's QE was estimated from equation~\ref{eqn:qe} using component level measurements
from manufacturers. 
The original LORRI ETC also relied on these component level measurements.
The system level QE was determined from ground calibration measurements \citep{morgan:2005}, and those measurements were used
to update the ETC.
Subsequently, in-flight measurements of absolute calibration standard reference stars (see the previous section) 
were used to further refine LORRI's absolute responsivity.
The final LORRI absolute calibration is essentially a hybrid product, with the wavelength dependence determined from the
ground calibration (when filters could be used to restrict the wavelengths sampled) and with the absolute sensitivity determined by
scaling the wavelength-dependent curve by whatever factor is needed to force a match between the predicted and observed LORRI signals for
observations of the absolute calibration standard stars.

\begin{figure}[htb!]
\includegraphics[keepaspectratio,width=\linewidth]{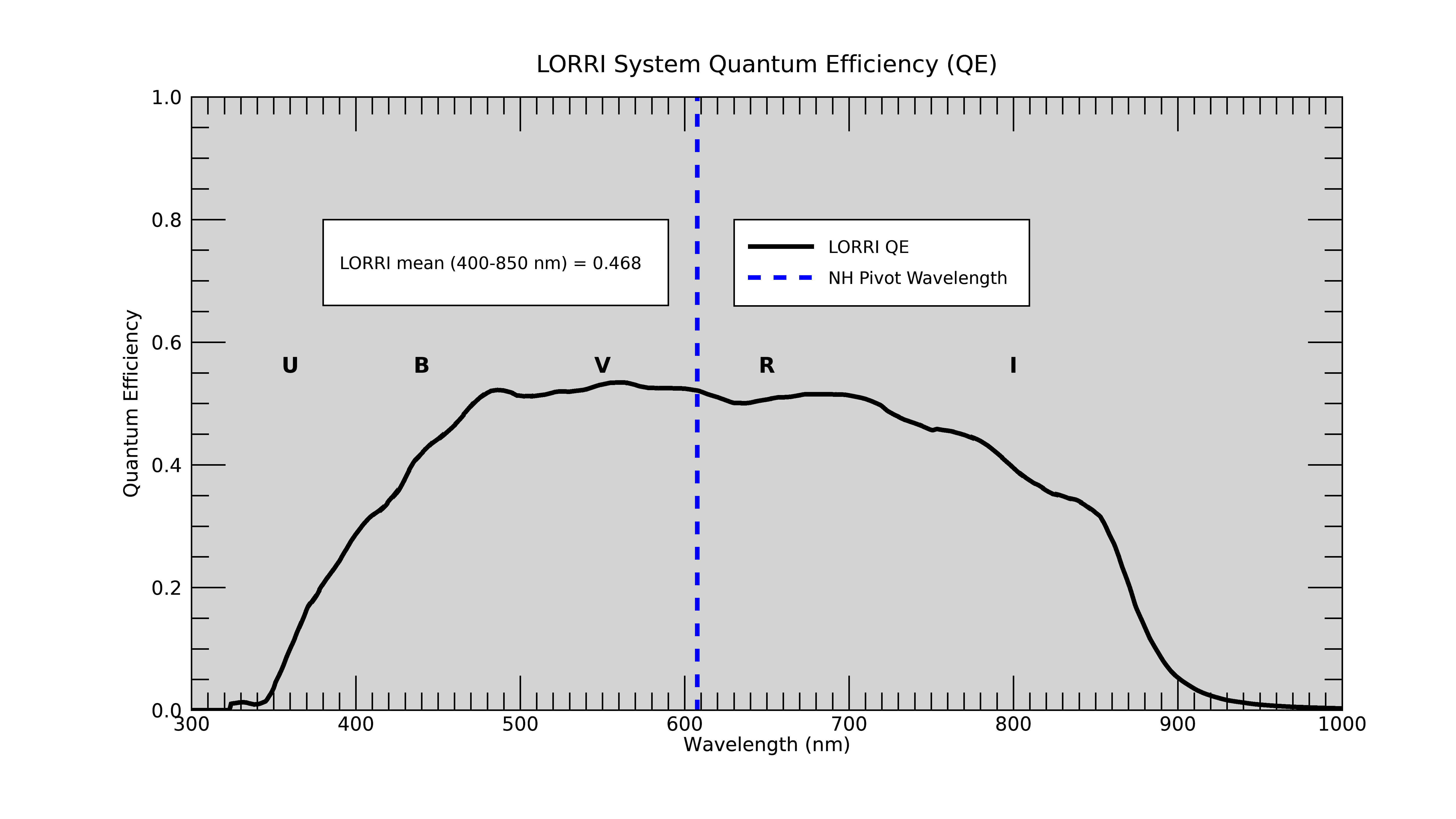}
\caption{LORRI's system quantum efficiency (QE) is plotted as a function of wavelength.
The mean QE over the wavelength range \mbox{$400-850$ nm} is $\sim$47\%.
The locations of the LORRI pivot wavelength and the standard visible photometric bands 
(in the Johnson-Landolt system) are also shown.
LORRI's pivot wavelength is located between the standard $V$ and $R$ band wavelengths.
}
\label{fig:qe}
\end{figure}

Figure~\ref{fig:qe} shows the LORRI system QE as a function of wavelength, as determined from the absolute calibration measurements
of the solar-type reference star HD~37962.
The LORRI system QE is approximately 50\% over much of the visible wavelength range (e.g., \mbox{480--700 nm}). 
LORRI is a panchromatic instrument, which means that its output signal is proportional to the integral over all wavelengths of the 
product of the QE and the target's SED.

\begin{figure}[htb!]
\includegraphics[keepaspectratio,width=\linewidth]{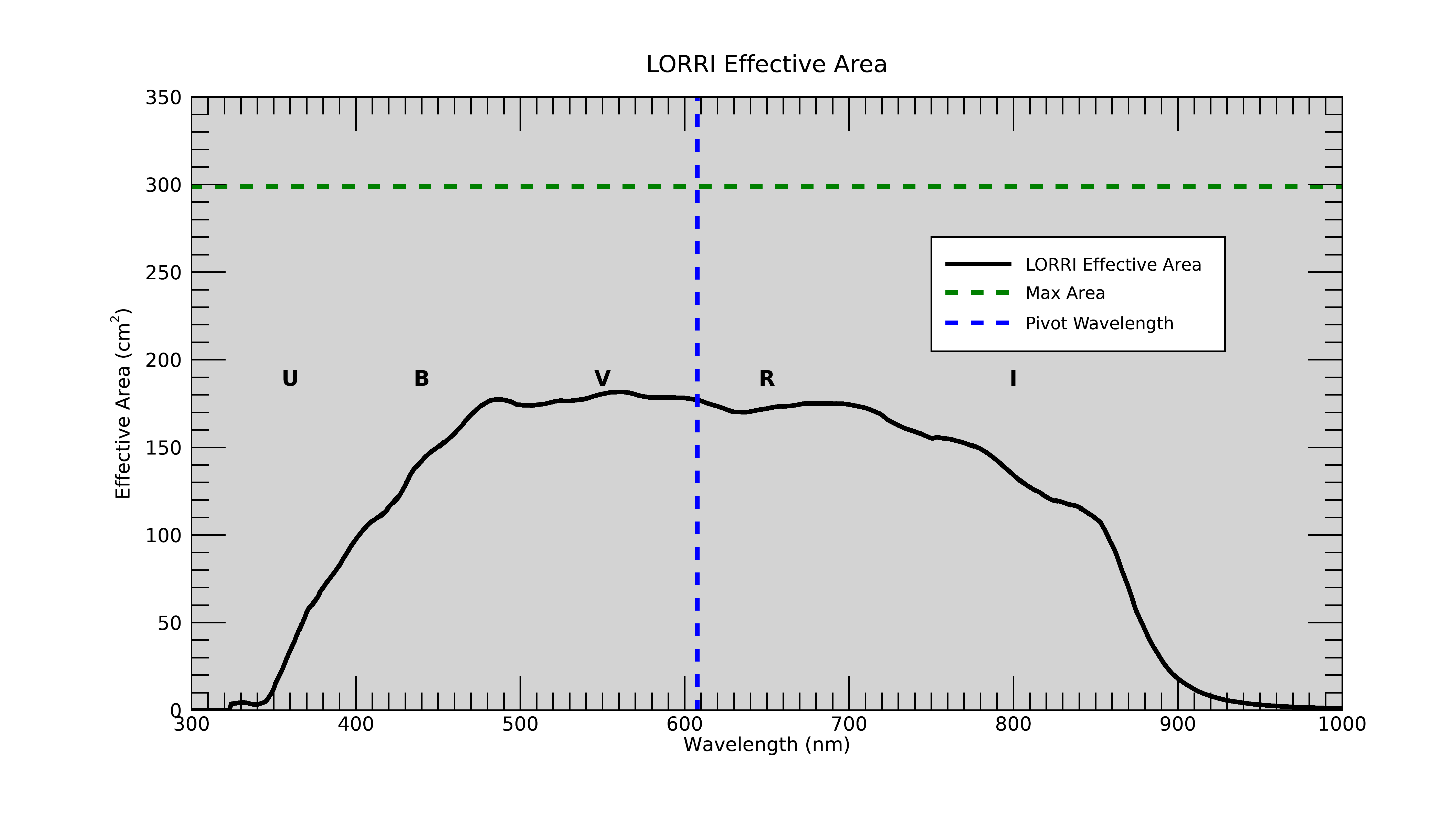}
\caption{LORRI's effective area is plotted as a function of wavelength.
Given the estimated LORRI obscuration of $\sim$11\%, the theoretical maximum
effective area is given by the green dashed line.
The locations of the LORRI pivot wavelength and the standard visible photometric bands are also shown.
}
\label{fig:aeff}
\end{figure}

Note that the commonly used quantity ``effective area'' ($\mathrm{A_{eff}}$) of the optical system is just the
area of LORRI's input aperture \mbox{(A $= \pi \ast 20.8^{2} / 4 = 339.8$ cm$^{2}$)} multiplied by the system QE.
Given the estimated LORRI obscuration of $\sim$11\% (i.e., $\mathrm{L_{obscure}} = 0.89$), the largest possible effective area for LORRI is
\mbox{$\sim$300  cm$^{2}$.}
Figure~\ref{fig:aeff} shows the LORRI effective area as a function of wavelength.

The LORRI absolute responsivity curve, which is used to calculate the photometry keywords, can be derived from the QE curve
using the following equation:
\begin{equation}
\mathrm{
R_{\lambda} = A_{eff} \ast \Omega \ast \lambda / gain / hc \\
}
\label{eqn:response}
\end{equation}

\noindent where:\\

\noindent $\mathrm{R_{\lambda}}$ is the responsivity ([DN s$^{-1}$ pixel$^{-1}$] / [ergs cm$^{-2}$ s$^{-1}$ sr$^{-1}$]) \\
$\mathrm{A_{eff}}$ is the effective area as defined above (cm$^{2}$) \\
$\mathrm{\Omega}$ is the solid angle of a single pixel (sr) \\
$\mathrm{\lambda}$ is the wavelength of interest (nm) \\
gain is the CCD gain (21 e DN$^{-1}$ for \onebyone format) \\
hc is the product of Planck's constant and the speed of light ($1.986 \times 10^{-16}$ J-nm) \\

\begin{figure}[htb!]
\includegraphics[keepaspectratio,width=\linewidth]{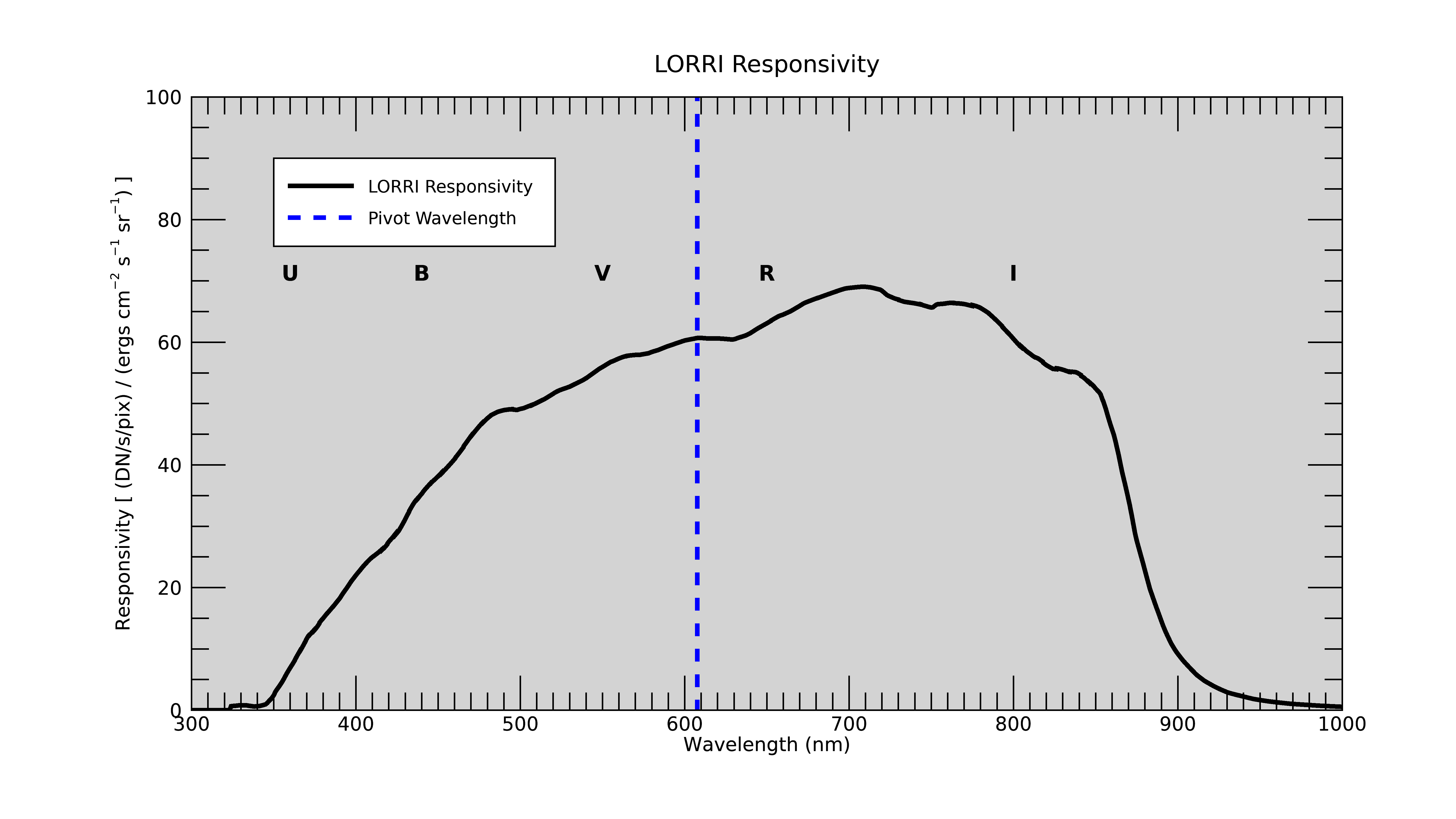}
\caption{LORRI's absolute responsivity is plotted as a function of wavelength.
The locations of the LORRI pivot wavelength and the standard visible photometric bands are also shown.
}
\label{fig:response}
\end{figure}

Figure~\ref{fig:response} shows the LORRI responsivity curve as a function of wavelength for \onebyone format, as
derived from the measurements of the absolute calibration standard star HD37962.
The responsivity curve for \fourbyfour format can be obtained by multiplying the \onebyone curve by the constant 17.1.

\clearpage

\section{Summary} \label{sec:summary}

LORRI's key design parameters are summarized in Table~\ref{tab:summary}.
LORRI has played a major role in the success of the \nh mission by
serving as the primary optical navigation camera, by providing the
highest resolution measurements of the mission's flyby targets,
and by performing high sensitivity observations of remote targets
at unique geometries.

\begin{deluxetable*}{ll}
\tablecaption{Summary of LORRI key parameters\label{tab:summary}}
\tablewidth{0pt}
\tablehead{
\colhead{Item} & \colhead{Description} 
}
\startdata
Optical Telescope Assembly (OTA) & L3H-SSG Ritchey-Chr\'{e}tien optical design with 3-element field flattener lens assembly \\
	& Silicon Carbide (SiC) structure, SiC mirrors coated with high reflectance dielectric \\
	& 20.8 cm primary mirror diameter, $\sim$11\% central obscuration \\
	& Focal length $=$ 261.908 cm \\
	& Plate scale $=$ 0.381813 $\mu$radians $\micron^{-1}$ $=$ 787\farcs546 mm$^{-1}$ \\
	& NH OTA in-flight operating temperature is approximately $-$70\arcdeg~C \\
	& No moving parts, except for once-open telescope cover mounted to spacecraft \\
Focal Plane Characteristics	& Teledyne-e2v 47-20 frame transfer CCD detector \\
		& CCD frame transfer time $\approx$12 ms \\
		& $1024 \times 1024$ optically active pixels, 13~$\micron$ square pixels \\
		& AR-coated, backside-thinned, backside-illuminated CCD \\
		& \onebyone and \fourbyfour (rebinned) output formats \\
		& IFOV (\onebyone format) $=$ 4.9636 $\mu$radians  $=$ $1\farcs$0231 (square)\\
		& FOV $=$ $5.0827$ milliradians $=$ $0\fdg$29122 (square) \\
		& CDS with 12-bit ADC \\
		& Full well $\approx$80,000 e (linear range) \\
		& Anti-blooming technology to mitigate the effects of saturated targets in the FOV \\
		& Electronics noise $\approx$24 e \\
		& Dynamic range $\approx$3500 (single image) \\
		& Gain: 21.0 e DN$^{-1}$ (\onebyone), 19.4 e DN$^{-1}$ (\fourbyfour) \\
		& Dark current $\leq$0.040 e s$^{-1}$ pixel$^{-1}$ (\onebyone at NH operating temperature of $-$81\arcdeg C) \\
		& Available exposure times: 0 ms to 64,967 ms at 1 ms spacings \\
		& 1 Hz maximum frame rate (minimum time between consecutive images is 1 s) \\
Wavelength Range	& Panchromatic (no filters) with $\sim$50\% peak QE \\
				& 435--870 nm at 50\% of peak QE \\
				& 360--910 nm at 10\% of peak QE \\
Photometric Accuracy	& $\sim$2\% (1$\sigma$) absolute for solar-type SED \\
					& $\leq$10\% (1$\sigma$) absolute for non-solar-type SEDs \\
					& $\leq$1\% (1$\sigma$) relative for SNR $\geq$100 \\
					& Stable sensitivity at $\sim$1\% level for $\geq$ 13 years \\
\enddata
\tablecomments{``NH'' stands for ``New Horizons'', ``CCD'' stands for ``Charge Coupled Device'',
``IFOV'' stands for ``Individual pixel Field of View'',
``FOV'' stands for ``Field of View'', and ``AR'' stands for ``Anti-Reflection''.
``CDS'' refers to ``Correlated Double Sampling'', ``ADC'' refers to analog-to-digital converter,
``QE'' stands for ``Quantum Efficiency'',``SED'' stands for ``Spectral Energy Distribution'',
and ``SNR'' stands for ``Signal-to-Noise Ratio''.
}
\end{deluxetable*}

Assuming that the wavelength variation of LORRI's sensitivity is accurately
described by the ground-based calibration, LORRI's absolute
sensitivity should be accurate to $\sim$2\% (1$\sigma$) for targets with solar-type SEDs.
The accuracy of the absolute calibration for targets with other SEDs
should be comparably good when employing synthetic photometry techniques,
which we do when deriving LORRI's photometry keywords.

LORRI's sensitivity and optical performance are essentially unchanged since the
launch of the \nh mission in January 2006, more than 13 years ago.
Although LORRI is a ``single string'' instrument, susceptible to a single point failure to
one of its critical components, its longevity is testimony to its simple, yet powerful, design.
Indeed, the next generation of LORRI is currently being built to serve similar functions
on NASA's \emph{Lucy} mission, which is scheduled to launch in October 2021 when it
will begin a program to perform the first flyby measurements of six Jovian Trojans \citep{lucy:2017}.

\acknowledgments
We thank the scientists, engineers, and managers at APL who participated in the design, construction, and
testing of LORRI prior to launch:
J. Boldt,
K. Cooper,
H. Darlington, 
M. Grey,
J. Hayes,
P. Hogue,
T. Magee,
E. Rossano,
and C. Schlemm.
We thank G. Rogers for his outstanding and innovative support of the spacecraft pointing operations,
which enabled the LORRI observing program.
We thank the \nh system engineering team (D. Kusnierkiewicz, C. Hersman, V. Mallder, G. Rogers) for its
excellent work in maintaining the health and safety of the spacecraft.
We thank S. Williams and A. Mick for their outstanding support of the \nh command and data handling system,
including their management of complex solid state recorder operations.
We thank the \nh Mission Operations team (especially A. Bowman, K. Whittenburg, and H. Hart) for its
expert implementation of the LORRI observational program.
We thank the \nh Science Operations team (E. Birath, A. Harch, D. Rose, and N. Martin) for its
expert scheduling of the LORRI observations and data downlink.
We thank M. Holdridge for his leadership during the Pluto and MU69 flyby campaigns.
We thank the personnel at NASA's Deep Space Network for their support
of communications with the  \nh spacecraft, including the downlinking 
of the mission engineering and science data.
We thank the \nh Project Managers, G. Fountain and H. Winters, for
their steadfast support of LORRI throughout the mission. 
We thank B. Carcich for developing a computational shortcut for the original desmear calculations.
We thank R. Bohlin for discussions regarding absolute calibration standards.
We thank the personnel at SSG Precision Optronics (now L3-Harris~SSG) who played major roles in building the LORRI OTA:
F. Azad,
K. E. Kosakowski,
and D. Sampath.

%



\software{Interactive Data Language (IDL), licensed by the Harris Corporation}
\clearpage

\bibliography{bibtex_refs_pasp_2019_v8_astroph}




\end{document}